\documentclass[%
 reprint,
nofootinbib,
 amsmath,amssymb,
 aps,
]{revtex4-2}

\usepackage{bm}
\usepackage{cleveref}
\usepackage{dcolumn}
\usepackage{textcomp}
\usepackage{gensymb}
\usepackage{graphicx}
\usepackage{physics}

\usepackage{changepage}   

\begin{document}

\preprint{APS/123-QED}

\title{Neutrino decoherence from quantum gravitational stochastic perturbations}


\author{Thomas Stuttard}
\author{Mikkel Jensen}%
\affiliation{%
Niels Bohr Institute, University of Copenhagen, DK-2100 Copenhagen, Denmark\
}%

\date{\today}

\begin{abstract}

Neutrinos undergoing stochastic perturbations as they propagate experience decoherence, damping neutrino oscillations over distance. Such perturbations may result from fluctuations in space-time itself if gravity is a quantum force, including interactions between neutrinos and virtual black holes. In this work we model the influence of heuristic neutrino-virtual black hole interaction scenarios on neutrino propagation and evaluate the resulting signals in astrophysical and atmospheric neutrinos. We derive decoherence operators representing these effects in the framework of open quantum systems, allowing experimental constraints on such systems to be connected to quantum gravitational effects. Finally, we consider the energy-dependence of such Planck scale physics at energies observed in current neutrino experiments, and show that sensitivity to Planck scale physics well below the `natural' expectation is achievable in certain scenarios.

\end{abstract}

\maketitle



\section{\label{sec:intro}Introduction}

The mixing between neutrino mass and flavor eigenstates produces the phenomena of neutrino oscillations, where a neutrino produced as one flavor may be detected some time later as another, and is well established experimentally~\cite{Fukuda:1998mi, Ahmad:2001an,Ahmad:2002jz}. This is a quantum superposition effect that is maintained over macroscopic distances due to the feeble interactions between neutrinos and matter, allowing the neutrino to propagate largely in isolation from its environment. Neutrino oscillations are thus generally considered to be \textit{coherent}, with the wavefunctions of two neutrinos of identical energy travelling along an identical path evolving identically.

If however there is weak (and as yet undetected) coupling between neutrinos and the environment in which they propagate, the neutrinos may experience stochastic perturbations to their wavefunctions as they travel, degrading or even completely destroying the coherence over large distances\footnote{This is a distinct phenomenon from wave packet decoherence~\cite{NUSSINOV1976201} produced via the separation of the neutrino mass states over long distances due to their differing masses.}. By contrast, the known modifications of neutrino oscillation probability due to the influence of matter\footnote{Non-Standard Interactions (NSI) also typically refers to coherent effects on neutrino propagation resulting from interactions between neutrinos and conventional matter via new forces.} such as the Mikheyev-Smirnov-Wolfenstein effect~\cite{Wolfenstein:1977ue,Mikheev:1986wj} and parametric resonances~\cite{Chizhov:1998ug,Akhmedov:1998ui} are the net result of the influence of many matter particles on a propagating neutrino, producing consistent effects for all traversing neutrinos and thus preserving coherence. 

A stochastic environment is a frequent prediction of quantum gravity models, with the postulated fluctuating nature of space-time at Planck scales (often referred to as \textit{space-time foam} or \textit{quantum foam}) perturbing the propagating neutrino~\cite{Hawking, PhysRev.97.511}. Searches for neutrino decoherence thus potentially afford us a rare window on Planck scale physics usually considered beyond the reach of current experiments.

The goal of this work is to investigate the characteristics of neutrino decoherence and other phenomena resulting from the influence of quantum gravity on neutrino propagation, focusing on the promising case of neutrino interactions with \textit{virtual black holes} produced by space-time fluctuations. To do so, we inject heuristic interaction scenarios into a software implementation of neutrino propagation to determine the resulting impact on neutrino flavor transitions. 

We then demonstrate how the derived phenomena can be represented in the framework of open quantum systems, which is commonly employed in neutrino decoherence phenomenology and experimental searches~\cite{PhysRevD.96.093009, Buoninfante:2020iyr, PhysRevD.99.075022, PhysRevLett.85.1166, OHLSSON2001159, Farzan:2008zv, Coloma:2018idr, Carpio:2018gum, Carpio:2017nui, Anchordoqui:2005gj, PhysRevD.95.113005, PhysRevD.91.053002, PhysRevD.76.033006, PhysRevLett.118.221801, GUZZO2016408, Morgan:2004vv, Abbasi:2009nfa, Gomes:2020muc, Ohlsson:2020gxx}. This framework is very general, making constraints on the parameters of the open quantum system difficult to physically interpret. This work therefore allows neutrino decoherence experimental constraints to be directly interpreted in terms of the underlying quantum gravitational phenomena considered here. We also consider the energy dependence of the physics tested, demonstrating that current experiments are sensitive to Planck scale physics well below the `natural' expectation in some scenarios, and compute the expected signal resulting from these effects in both astrophysical and atmospheric neutrinos. 


\section{\label{sec:decoh_from_perturbations}Decoherence from stochastic perturbations}

Neutrinos propagate as mass states, and the evolution of a relativistic neutrino mass state can be represented as a plane wave:

\begin{equation}
\label{eqn:plane_wave}
\ket{\nu_{j}(L)} = \exp{ -i \frac{ m_j^2 L }{ 2 E } } \ket{\nu_{j}(0)} ,
\end{equation}

\noindent where $\ket{\nu_{j}}$ is the neutrino mass state $j$ ($j=1,2,3$ in the $3v$ paradigm) of mass $m_j$, with $E$ being the neutrino energy and $L$ the distance travelled. 

Neutrino mass states can be propagated according to \Cref{eqn:plane_wave}, with the oscillation probability after a given distance being determined by rotating the current state to the neutrino flavor basis, as defined by the Pontecorvo-Maki-Nakagawa-Sakata (PMNS) mixing matrix~\cite{Pontecorvo:1957qd, Maki:1962mu}, and projecting onto the desired final flavor state according to: 

\begin{equation}
\label{eqn:osc_prob_projection}
P( \nu_\alpha \rightarrow \nu_\beta ) = |\braket{\nu_{\beta}(L)}{\nu_{\alpha}(0)}|^2 ,
\end{equation}

\noindent where $\alpha,\beta$ represent flavor indices ($e,\mu,\tau$ in the $3v$ paradigm).

Decoherence can result from stochastic perturbations to the mass states as they propagate, for example from perturbations to the phase of one or more of the neutrino mass states. Such a phase perturbation can in included in \Cref{eqn:plane_wave} as an additional term, $\delta \phi_j(L)$: 

\begin{equation}
\label{eqn:plane_wave_perturbed}
\ket{\nu_{j}(L)} = \exp{ -i \left( \frac{ m_j^2 L }{ 2 E } + \delta \phi_j(L) \right) } \ket{\nu_{j}(0)} .
\end{equation}

\begin{figure}[htp]
    \centering
    \includegraphics[trim=0.5cm 0.0cm 0.5cm 0.0cm, clip=true, width=\linewidth]{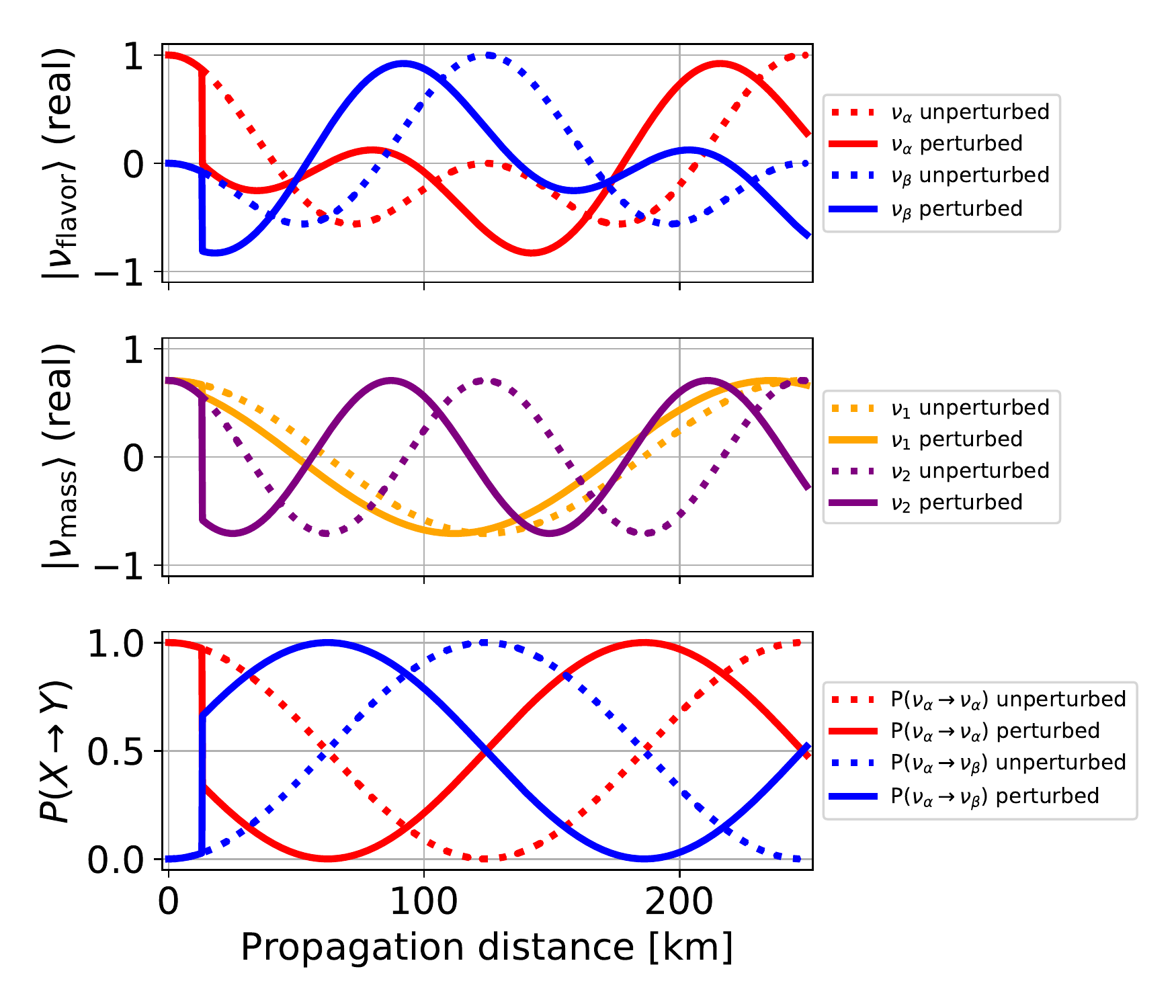}
    \caption{Impact of a perturbation to the phase of the propagating neutrino states, where the neutrino is initially in a pure $\nu_\alpha$ state. The neutrino mass state phase is perturbed at a randomized distance, in this example at $L\sim$ 10\,km. The parameters defined in \Cref{table:2nu_params} are used, and the mixing angle, $\theta = 45 \degree$. Only the real components of the flavor/mass states are shown for clarity. For comparison, the dotted lines indicate the state evolution in the case of no perturbation.}
    \label{fig:toy_model_perturb}
\end{figure}

\begin{figure}[htp]
    \centering
    \includegraphics[width=\linewidth]{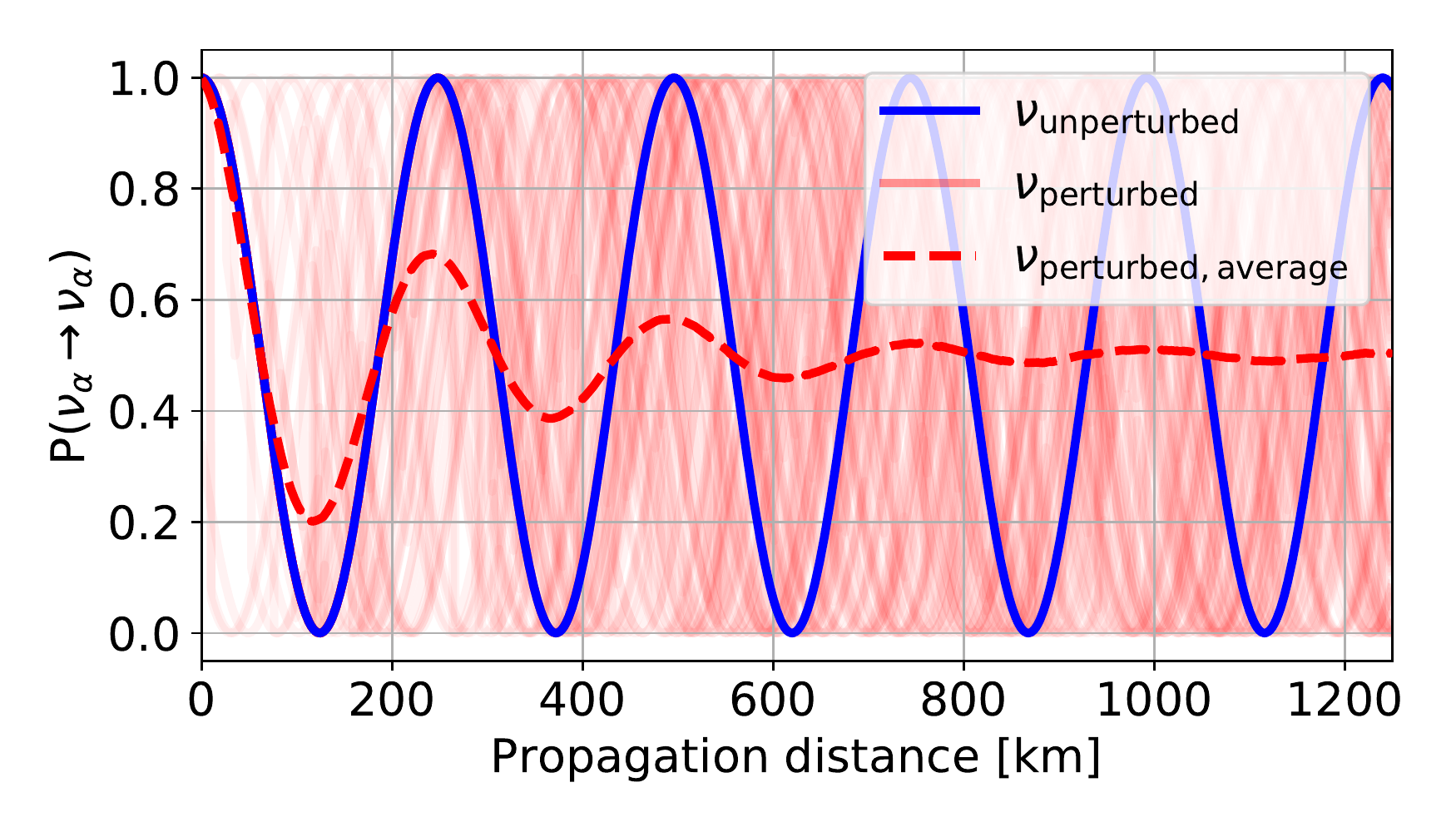}
    \caption{Oscillation (survival) probability for many perturbed neutrinos, for the same system shown in \Cref{fig:toy_model_perturb}. Each light red line shows the probability for a single neutrino undergoing stochastic phase perturbation, whilst the red dashed line shows the average oscillation probability of the whole population. The blue line shows the oscillation probability in the absence of perturbations. The point in space at which the perturbation occurs is randomly chosen according to a mean free path of 250\,km.}
    \label{fig:toy_model_net_decoh}
\end{figure}

A example of the impact of such a phase perturbation on the propagating neutrino states  is shown in \Cref{fig:toy_model_perturb}, where the perturbation to each mass state\footnote{Note that the phase perturbations must differ for each mass state, as neutrino oscillations are invariant to a global phase shift.} is injected at a random distance, with the perturbation strength randomly sampled from the interval $[0, 2\pi]$. Following a perturbation, the mass states continue to evolve as before (with the same frequency and amplitude), but with a shifted phase. When many neutrinos are considered, the probability of each neutrino having undergone a perturbation increases with distance, and the population becomes increasingly out of phase. This results in damping of the average oscillation probability with increasing distance, as shown in \Cref{fig:toy_model_net_decoh}, ultimately resulting in a total loss of coherence at large distances. We will show in this work that this intuitive picture of perturbed phases, as well as a range of other types of stochastic perturbations, are completely captured by the open quantum system formalism of decoherence.

Note that even after full decoherence, flavor transitions are still taking place in \Cref{fig:toy_model_net_decoh} (in this case $P(\nu_\alpha \rightarrow \nu_\alpha) \sim 0.5$ due to the maximal value of $\theta$ used), but the probability ceases to vary with time/distance. We will later show how the fully decohered behaviour of the system varies for different forms of perturbation.

\begin{table}
 \begin{tabular}{|c| c|} 
 \hline
 Parameter & Value \\
 \hline
 \hline
 \# states & 2 \\ \hline
 $m_1$ & 0.1\,eV \\ \hline
 $m_2$ & $\sqrt{2} m_1$ \\ \hline
 E & 1\,GeV \\ \hline
 Initial flavor & $\nu_\alpha$ \\ \hline 
\end{tabular}
\caption{Parameters used for the propagating 2$\nu$ system. The mass states are labelled $0,1$ and the flavor states $\alpha,\beta$. The parameter values are chosen to produce clear demonstrations of the behaviour, rather than to represent realistic neutrino parameters.}
\label{table:2nu_params}
\end{table}

\subsection{\label{sec:pertubation_types}Neutrino perturbations from quantum gravity}

If gravity is a quantum force subject to the uncertainty principle, it is hypothesised that the space-time itself fluctuates at the Planck scale~\cite{Hawking, PhysRev.97.511, ELLIS199237}, often referred to as \textit{space-time foam}~\cite{misner1973gravitation,ng2010holographic}. Such fluctuations in space-time curvature imply fluctuations in the travel distance/time between two points (e.g.\@ the space-time \textit{metric}), a phenomenon known as \textit{lightcone fluctuations}~\cite{PauliLightcone, Ford_1995,Yu_2009}. In such a scenario, one might expect fluctuations in the time taken for neutrinos to propagate from a source to a detector, thus varying the neutrino mass state wavefunction at the point of detection. How strongly a particle is influenced by Planck scale fluctuations would likely depend on the particle's energy relative to the Planck mass, e.g.\@ how clearly it would `see' features at this scale.

At the extreme, fluctuations in the space-time foam of sufficient magnitude could collapse to form black holes of Planck length scale, which would almost immediately evaporate (at Planck time scales). These \textit{virtual black holes} (VBH) are analogous to the virtual electron-positron pairs that form the phenomenon of \textit{vacuum polarization} in quantum electrodynamics (QED). Neutrinos encountering these black holes may experience loss of quantum information or other strong perturbations. As one example, a neutrino might be absorbed by the black hole, with the black hole subsequently evaporating/decaying to produce new particles altogether, conserving only energy, charge and angular momentum (as per the \textit{no hair theorem}~\cite{BARROW198712}) but not baryon or lepton number. Such processes have been proposed as a source of proton decay, where the constituent quarks of the otherwise stable proton are absorbed by the VBH and re-emitted as other particles~\cite{ProtonDecay,Alsaleh_2017}. Heuristically, the neutrino may be viewed as being stochastically absorbed and (possibly) re-emitted by these VBH encounters during propagation, with this stochasticity potentially resulting in decoherence. 

Unlike lightcone fluctuations where significant effects would be expected to accumulate over very long propagation distances~\cite{ng2010holographic,Perlman_2015}, $\nu$-VBH interactions could produce significant effects over more modest distances provided they occur with sufficient frequency due to the potentially strong perturbation experienced by the neutrino during even a single VBH encounter. We focus on this case in this work.

In addition to the quantum effects considered in this work, neutrino decoherence resulting from classical gravitation has also been studied~\cite{CHATELAIN2020135150, PhysRevD.100.096014}. More mundane sources of decoherence in neutrino oscillation measurements have also previously been identified that must not be confused with the effects of quantum gravity or other new physics. For example, the spatial extent of the neutrino source or other variations in the source-detector distance in a neutrino experiment can produce decoherence effects, and occurs for instance due to variations in the height of cosmic ray air showers producing atmospheric neutrinos. Additionally, conventional neutrino-matter effects feature some degree of decoherence, for example from sub-structure in the Earth's internal density distribution~\cite{MatterDensityFluctuation} or non-forward scattering~\cite{2002.08315}. Detector resolution also introduces a form of decoherence into measurements~\cite{OHLSSON2001159}. The characteristics of any detected neutrino decoherence effects must therefore be studied carefully to try to separate different scenarios, where the strong energy-dependence that might be expected to result from Planck scale physics in particular may prove a useful handle for separating quantum gravity effects.

\subsection{\label{sec:modelling_nuVBH_interactions}Modelling $\nu$-VBH interactions}

We now evaluate the influence of $\nu$-VBH interactions on neutrino propagation and oscillations. Given the absence of an accepted model of quantum gravity, we test a series of heuristic scenarios designed to capture the potential microphysics of these interactions. Four potential cases for the nature of the interaction/perturbation are tested:
\vspace{1.5mm} \\
\textbf{Mass state selected:} The interaction selects a single neutrino mass state. The state is selected democratically, i.e. with equal probability for any state.
\vspace{1.5mm} \\
\textbf{Flavor state selected:} The interaction selects a single definite flavor state, selected democratically as described for the mass state case. Lepton number is potentially violated in the interaction. 
\vspace{1.5mm} \\
\textbf{Large phase perturbation:} The neutrino experiences large (but otherwise unspecified) perturbations to its mass state phases, which are essentially randomized. 
\vspace{1.5mm} \\
\textbf{Neutrino loss:} The neutrino is lost in the interaction and not observed. This could result from the neutrino being swallowed by the black hole and either lost or re-emitted via Hawking radiation as another (non-detected) particle type due to the lack of global symmetry conservation. An alternative picture would be that the outgoing neutrino is simply re-emitted in a different direction and thus not observed (particularly for a distant source). This is the only non-unitary case tested (where information is lost to the environment), and is phenomenologically similar to neutrino decay scenarios~\cite{LINDNER2001326,BERRYMAN201574} (although likely with differing energy-dependence).

\begin{figure}
    \centering
    \includegraphics[trim=0.1cm 0.0cm 0.1cm 0.0cm, clip=true, width=\linewidth]{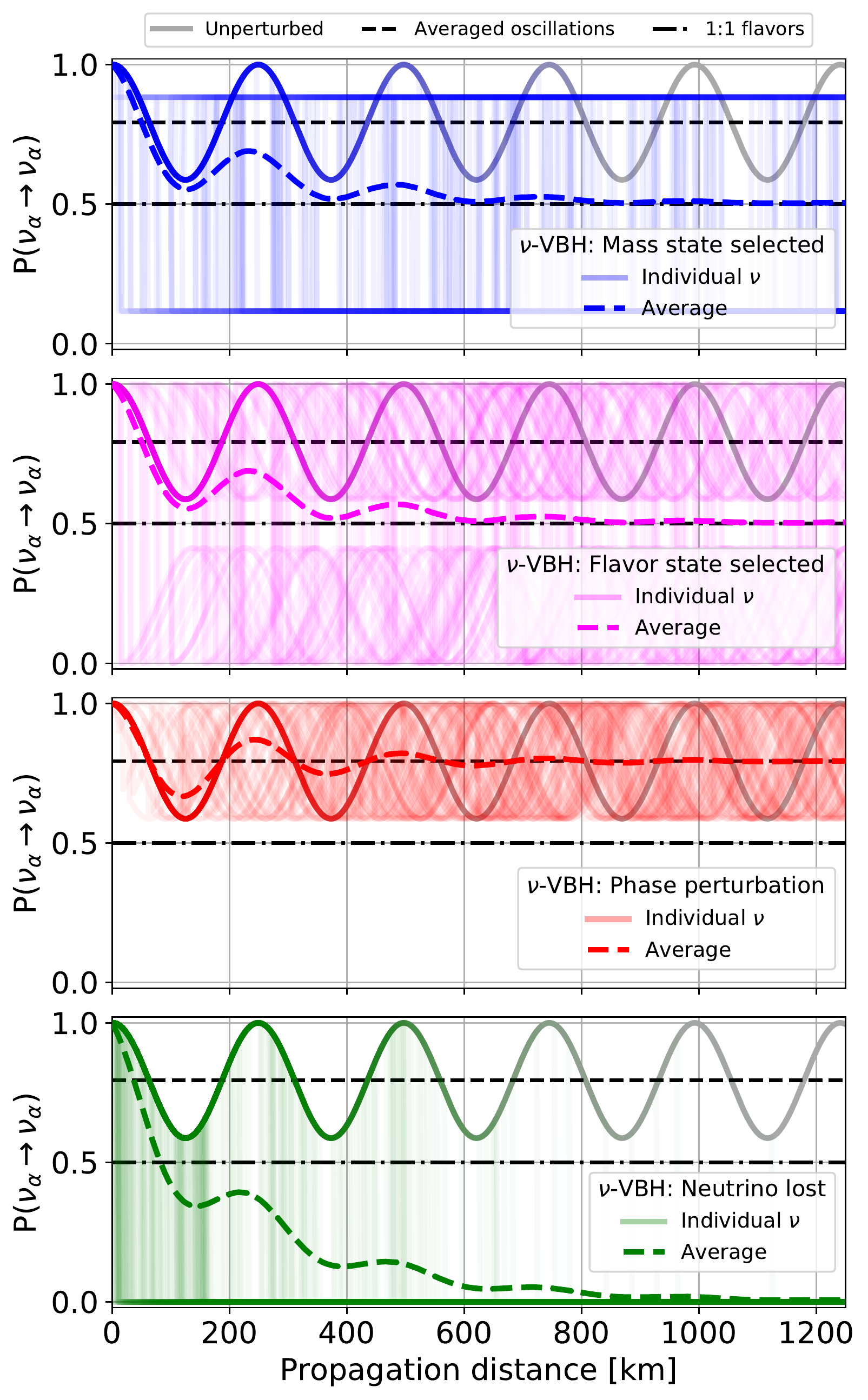}
    \caption{Neutrino flavor transition probability resulting from various $\nu$-VBH interaction scenarios. Both individual neutrinos (translucent coloured lines) and the average behaviour of the ensemble (opaque dashed coloured lines) are shown (note that only the ensemble behaviour is observable). Note that in some regions of the plots, many overlaid translucent lines result in solid coloured lines. The neutrinos are initially in a pure $\nu_\alpha$ state. A two flavor system is shown using the toy parameters in \Cref{table:2nu_params}, with a non-maximal mixing angle $\theta=20\degree$. The interaction mean free path is 250\,km. }
    \label{fig:toy_model_vVBH_scenarios}
\end{figure}

To determine the influence of these interactions on neutrino flavor transitions, we propagate neutrinos as described in Section \ref{sec:decoh_from_perturbations} and inject the interactions described above at randomised distances according to an interaction \textit{mean free path}. This mean free path is the lone free parameter of the system, and is the product of the VBH number density along the neutrino travel path and the interaction cross section. For each interaction scenario we propagate many individual neutrinos and compute the average behaviour of the neutrino ensemble. The resulting neutrino survival probabilities versus distance are shown in \Cref{fig:toy_model_vVBH_scenarios}. As in Section \ref{sec:decoh_from_perturbations}, a 2 flavor system is shown with toy model parameters chosen for clarity. In particular, a non-maximal mixing angle $\theta$ is used. A three flavor system with realistic parameters is shown later in Section \ref{sec:oqs_vs_perturbation}.

In all scenarios, we observe the damping of the average oscillation probability of the neutrino ensemble that is characteristic of neutrino decoherence, ultimately resulting in a distance-independent flavor transition probability at large distances. The main difference between the cases is the behavior at large distances, e.g.\@ when the neutrino population has fully lost coherence or when all neutrinos have been lost (these differences are discussed in more detail below). These differences could potentially be used to discriminate between the scenarios in the event of an experimental signal being observed. 

The rate of damping is identical in all cases, as the perturbed final states are independent of the initial states and thus the perturbations totally eliminate coherence for that neutrino. The damping rate is thus purely defined by the interaction mean free path, which controls the fraction of the neutrino ensemble that have experienced one (or more) interactions after a given distance. All cases show clear deviation from standard (unperturbed) oscillations, and can therefore be searched for experimentally.

We now discuss the individual scenarios in more detail. For the `neutrino loss' case, the neutrinos follow standard oscillation behaviour until they undergo an interaction, at which point the neutrino is lost and the transition probability (to any flavor/state) immediately drops to 0. The long distance behaviour of both individual neutrinos and the ensemble is thus $P(\nu_X \rightarrow \nu_X ) = 0$, where $\nu_X$ represents any neutrino flavor. We note that although this case shares the phenomenological characteristics of neutrino decoherence (and we will later see it can be expressed in the same mathematical framework), it is not strictly a form of decoherence as neutrinos are removed from the system, rather than losing coherence with the population.

In the `phase perturbation' scenario, once the neutrino undergoes an interaction it becomes out of phase with the neutrino population, but continues to oscillate. At large distance, eventually all neutrinos have experienced an interaction and coherence is totally lost in the ensemble, resulting in an averaging of the oscillation behaviour of the system. The long distance flavor transition probability is thus the averaged oscillation probability, $P(\nu_\alpha \rightarrow \nu_\beta ) = \sum_j |U_{\alpha j}|^2 |U_{\beta j}|^2$ (where $U$ is the PMNS mixing matrix), and is mixing angle dependent.

In both the `mass state selection' and `flavor state selection' scenarios, the result of an interaction is that the system is forced to align with a particular state vector, in the mass or flavor basis respectively. This can be seen in the upper two panels of \Cref{fig:toy_model_vVBH_scenarios}, where the individual neutrinos separate into two distinct populations corresponding to the two mass/flavor states in the system (this would be three populations in a three neutrino system). Neutrinos may switch between populations as they continue to propagate and potentially encounter further VBHs. The long distance behaviour of the ensemble in both these cases is equal numbers of neutrinos in each population, resulting in an average transition probability of $P(\nu_X \rightarrow \nu_X ) = \frac{1}{N}$ (where $N$ is the number of neutrino states considered, in this case $N=2$), independent of the neutrino mixing angle(s). The behaviour of the ensemble is identical regardless of whether a mass or flavor state is selected in the interaction, and thus these cases are indistinguishable through a neutrino oscillation measurement. For the flavor state case, individual neutrinos continue to oscillate following the interaction (although out of phase with each other since the interaction takes place at an random phase in the wavefunction evolution). For the mass state case however, the selection of a single mass state destroys the superposition effect caused by the co-evolution of multiple mass states that causes the time-dependent oscillatory characteristics of the flavor transitions. Flavor transitions are still possible for individual neutrinos though, e.g. $P(\nu_X \rightarrow \nu_X ) \neq 0,1$, due to the mixing of mass and flavor states, but in a time-independent manner.

An interesting observation from this study is that the `phase perturbation' case can appear similar or even identical to the other cases in certain $\nu_\alpha \rightarrow \nu_\beta$ channels for particular mixing angles. For example, in a two state system with maximal mixing ($\theta=45\degree$), the `phase perturbation' scenario produces identical ensemble damping effects to the `mass/flavor state selection' scenarios (with large distance behaviour of $P(\nu_\alpha \rightarrow \nu_\beta) = 0.5$). This is a good approximation of e.g. high-energy atmospheric neutrino oscillations, limiting the distinguishability of these scenarios in such cases.

Now we have demonstrated the resulting signal for four $\nu$-VBH interaction scenarios by injecting perturbations into a software model of neutrino propagation, we will now look to represent this physics in the open quantum system formalism often used to represent neutrino decoherence.


\section{Decoherence in open quantum systems}
\label{sec:open_quantum_system_decoh}

A neutrino coupled to its environment can be treated using an open quantum system formalism. Although the neutrino may be produced in a known state, the random nature of the perturbations discussed in this work mean that the observer becomes increasingly ignorant of the neutrino's state as it propagates. The state can then only be expressed as an ensemble of possible states, each with an associated probability, known as a mixed quantum state. In the language of open quantum systems, decoherence is thus the transition from an initial pure quantum state to a mixed quantum state.

Mixed (and pure) quantum states can be mathematically expressed using the density matrix formalism, where the density matrix, $\rho$, for a system of $j$ states of probability $p_j$ is given by: 

\begin{equation}
\label{eqn:density_matrix}
\rho = \sum_j p_j \ket{\psi_j} \bra{\psi_j} .
\end{equation}

The density matrix for a pure quantum state is thus $\rho = \ket{\psi} \bra{\psi}$. Density matrices are suitable for describing both the state of a single neutrino and an ensemble.

The time evolution of an open quantum system experiencing decoherence is given by the Lindblad master equation~\cite{lindblad1976}:

\begin{equation}
\label{eqn:decoh_master}
\dot{\rho} = -i[H,\rho] - \mathcal{D}[\rho] ,
\end{equation}

\noindent where $H$ is the Hamiltonian of the system and $\mathcal{D}[\rho]$ is an operator defining decoherence in the system. Conventional coherent matter effects appear in the Hamiltonian. The neutrino flavor transition probability can be determined by:

\begin{equation}
\label{eqn:density_matrix_transition_prob}
P(\nu_\alpha \rightarrow \nu_\beta) = \mathrm{Tr}[\rho_\alpha(t)\rho_\beta(0)] .
\end{equation}

The form of $\mathcal{D}[\rho]$ is dependent on the underlying physics producing the decoherence effect. A generalised form of $\mathcal{D}[\rho]$ is~\cite{Benatti_2000,PhysRevLett.85.1166,gago2002study}:

\begin{equation}
\label{eqn:lindblad_decoh_operator}
\mathcal{D}[\rho] = - \frac{1}{2} \sum_k^{N^2-1} \left( [ V_k, \rho V_k^\dagger ] + ( [ V_k \rho, V_k^\dagger ] \right) ,
\end{equation}

\noindent where $N$ is the dimensionality of the $SU(N)$ Hilbert space defining the system ($SU(3)$ for a system with 3 neutrino flavors) and $V_k$ are $N \times N$ complex matrices.

The general $\mathcal{D}[\rho]$ form shown in \Cref{eqn:lindblad_decoh_operator} in principle allows model-independent decoherence searches to be performed, but in practise contains far too many free parameters to be realistically testable. Studies have considered only a small number of non-zero parameters, either selected for simplicity or to target some particular physics case. Here, we seek to reproduce the effects of the $\nu$-VBH interaction scenarios investigated in Section \ref{sec:modelling_nuVBH_interactions} using this open quantum system formalism.

It is common to expand the $\mathcal{D}[\rho]$ operator in terms of the basis vectors, $b_\mu$ of the $SU(N)$ space defining the system~\cite{Benatti_2000,gago2002study,PhysRevD.99.075022,Buoninfante:2020iyr}:

\begin{equation}
\mathcal{D}[\rho] = c_\mu b^\mu ,
\label{eqn:su3_expansion}
\end{equation}

\noindent where $c_\mu \equiv(\mathcal{D}[\rho])_\mu$, e.g.\@ the $\mu$'th coefficient of the $\mathcal{D}[\rho]$ expansion. The Einstein summation convention is used here. 

For a 3 neutrino system, $b_\mu$ are given by the $SU(3)$ generators, the \textit{Gell-Mann matrices}, and the identity matrix:

\begin{equation}
\begin{array}{ccc}

b_0 = \begin{pmatrix} 1 & 0 & 0\\0 & 1 & 0\\0 & 0 & 1 \end{pmatrix},
b_1 = \begin{pmatrix} 0 & 1 & 0\\1 & 0 & 0\\0 & 0 & 0 \end{pmatrix},
b_2 = \begin{pmatrix} 0 & -i & 0\\i & 0 & 0\\0 & 0 & 0 \end{pmatrix}, \\ 
b_3 = \begin{pmatrix} 1 & 0 & 0\\0 & -1 & 0\\0 & 0 & 0 \end{pmatrix}, 
b_4 = \begin{pmatrix} 0 & 0 & 1\\0 & 0 & 0\\1 & 0 & 0 \end{pmatrix}, 
b_5 = \begin{pmatrix} 0 & 0 & -i\\0 & 0 & 0\\i & 0 & 0 \end{pmatrix}, \\ 
b_6 = \begin{pmatrix} 0 & 0 & 0\\0 & 0 & 1\\0 & 1 & 0\end{pmatrix}, 
b_7 = \begin{pmatrix} 0 & 0 & 0\\0 & 0 & -i\\0 & i & 0\end{pmatrix}, \notag
b_8 = \frac{1}{\sqrt{3}} \begin{pmatrix} 1 & 0 & 0\\0 & 1 & 0\\0 & 0 & -2\end{pmatrix}.

\label{eqn:su3_basis}
\end{array}
\end{equation}

To define the free parameters, we can express the decoherence operator as:

\begin{equation}
\mathcal{D}[\rho] = (D_{\mu\nu} \rho^\nu) b^\mu ,
\label{eqn:decoh_operator_sun}
\end{equation}

\noindent where $\rho^\nu$ are the coefficients of the system's density matrix expanded in the $SU(N)$ basis (e.g.\@ $\rho = \rho_\mu b^\mu$), and $D_{\mu\nu}$ are the elements of a ($N^2 \times N^2$) matrix whose elements are the free parameters of the system ($D_{\mu\nu} \rho^\nu = c_\mu$ as defined in \Cref{eqn:su3_expansion}). For a 3 neutrino system, $D$ is defined as\footnote{Care must be taken when comparing $D$ between different studies, as the elements depend on the choice (and order) of basis vectors in which they are defined.}:

\begin{equation}
\label{eqn:decoherence_gamma_matrix}
D = \begin{pmatrix}
\Gamma_{0} & \beta_{01} & \beta_{02} & \beta_{03} & \beta_{04} & \beta_{05} & \beta_{06} & \beta_{07} & \beta_{08} \\ 
\beta_{01} & \Gamma_{1} & \beta_{12} & \beta_{13} & \beta_{14} & \beta_{15} & \beta_{16} & \beta_{17} & \beta_{18} \\ 
\beta_{02} & \beta_{12} & \Gamma_{2} & \beta_{23} & \beta_{24} & \beta_{25} & \beta_{26} & \beta_{27} & \beta_{28} \\ 
\beta_{03} & \beta_{13} & \beta_{23} & \Gamma_{3} & \beta_{34} & \beta_{35} & \beta_{36} & \beta_{37} & \beta_{38} \\ 
\beta_{04} & \beta_{14} & \beta_{24} & \beta_{34} & \Gamma_{4} & \beta_{45} & \beta_{46} & \beta_{47} & \beta_{48} \\ 
\beta_{05} & \beta_{15} & \beta_{25} & \beta_{35} & \beta_{45} & \Gamma_{5} & \beta_{56} & \beta_{57} & \beta_{58} \\ 
\beta_{06} & \beta_{16} & \beta_{26} & \beta_{36} & \beta_{46} & \beta_{56} & \Gamma_{6} & \beta_{67} & \beta_{68} \\ 
\beta_{07} & \beta_{17} & \beta_{27} & \beta_{37} & \beta_{47} & \beta_{57} & \beta_{67} & \Gamma_{7} & \beta_{78} \\ 
\beta_{08} & \beta_{18} & \beta_{28} & \beta_{38} & \beta_{48} & \beta_{58} & \beta_{68} & \beta_{78} & \Gamma_{8} \\ 
\end{pmatrix} ,
\end{equation}

\noindent where the diagonal parameters are indicated by $\Gamma_{\mu}$ and the off-diagonal elements by $\beta_{\mu\nu}$ (all are real scalars).

Although there are a large number of free parameters in $D$,  fairly general conditions such as probability and energy conservation can be imposed to reduce this matrix~\cite{Benatti_2000,PhysRevLett.85.1166,gago2002study}. For example, elements in the 0'th row and column of $D$ (those corresponding to the identity matrix) must be zero for a unitary system where no probability is lost from the neutrino to the environment~\cite{Buoninfante:2020iyr}, and thus are often omitted. Ultimately, the parameter values are chosen to represent the particular physics case of interest, or in some works a minimal set of non-zero parameters is (often somewhat arbitrarily) chosen to allow the formalism to be tested against experimental data.


\subsection{\label{sec:oqs_vs_perturbation}Representing $\nu$-VBH interactions in the open quantum system formalism}

Now that we have a formalism for characterizing the influence of the environment  on neutrino propagation within the context of an open quantum system, we seek to represent the $\nu$-VBH interaction scenarios (specifically the average behaviour of the ensemble) examined in Section \ref{sec:decoh_from_perturbations} in this framework by choosing appropriate forms for $D$.

All $\nu$-VBH interaction scenarios tested in this work produce exponential damping behavior of the form $e^{-\alpha L}$, where $\alpha$ represents a damping constant. Inspection of \Cref{eqn:decoh_master} therefore implies $\mathcal{D}[\rho]$ terms of the general form $\alpha \rho$. The damping constants will be specified in the $D$ matrix.

Where the scenarios differ is the large distance flavor transition probability they tend to after full decoherence or neutrino loss. The `mass state selected' and `flavor state selected' cases produce identical results for the ensemble, and thus can be represented by a single $D$ matrix. All cases ultimately depend on a single free parameter, the $\nu$-VBH interaction mean free path, and thus we also seek a single free parameter in the open quantum system description for each case. The following three $D$ matrices reproduce the $\nu$-VBH interaction cases in this work:

\begin{equation}
\label{eqn:randomized_flavor_D}
D_{\rm{state\:selected}} = \begin{pmatrix}
0 & 0 & 0 & 0 & 0 & 0 & 0 & 0 & 0 \\ 
0 & \Gamma & 0 & 0 & 0 & 0 & 0 & 0 & 0 \\ 
0 & 0 & \Gamma & 0 & 0 & 0 & 0 & 0 & 0 \\ 
0 & 0 & 0 & \Gamma & 0 & 0 & 0 & 0 & 0 \\ 
0 & 0 & 0 & 0 & \Gamma & 0 & 0 & 0 & 0 \\ 
0 & 0 & 0 & 0 & 0 & \Gamma & 0 & 0 & 0 \\ 
0 & 0 & 0 & 0 & 0 & 0 & \Gamma & 0 & 0 \\ 
0 & 0 & 0 & 0 & 0 & 0 & 0 & \Gamma & 0 \\ 
0 & 0 & 0 & 0 & 0 & 0 & 0 & 0 & \Gamma \\ 
\end{pmatrix} ,
\end{equation}

\begin{equation}
\label{eqn:randomized_phase_D}
D_{\rm{phase\:perturbation}} = \begin{pmatrix}
0 & 0 & 0 & 0 & 0 & 0 & 0 & 0 & 0 \\ 
0 & \Gamma & 0 & 0 & 0 & 0 & 0 & 0 & 0 \\ 
0 & 0 & \Gamma & 0 & 0 & 0 & 0 & 0 & 0 \\ 
0 & 0 & 0 & 0 & 0 & 0 & 0 & 0 & 0 \\ 
0 & 0 & 0 & 0 & \Gamma & 0 & 0 & 0 & 0 \\ 
0 & 0 & 0 & 0 & 0 & \Gamma & 0 & 0 & 0 \\ 
0 & 0 & 0 & 0 & 0 & 0 & \Gamma & 0 & 0 \\ 
0 & 0 & 0 & 0 & 0 & 0 & 0 & \Gamma & 0 \\ 
0 & 0 & 0 & 0 & 0 & 0 & 0 & 0 & 0 \\ 
\end{pmatrix} ,
\end{equation}

\begin{equation}
\label{eqn:neutrino_loss_D}
D_{\rm{neutrino\:loss}} = \begin{pmatrix}
\Gamma & 0 & 0 & 0 & 0 & 0 & 0 & 0 & 0 \\ 
0 & \Gamma & 0 & 0 & 0 & 0 & 0 & 0 & 0 \\ 
0 & 0 & \Gamma & 0 & 0 & 0 & 0 & 0 & 0 \\ 
0 & 0 & 0 & \Gamma & 0 & 0 & 0 & 0 & 0 \\ 
0 & 0 & 0 & 0 & \Gamma & 0 & 0 & 0 & 0 \\ 
0 & 0 & 0 & 0 & 0 & \Gamma & 0 & 0 & 0 \\ 
0 & 0 & 0 & 0 & 0 & 0 & \Gamma & 0 & 0 \\ 
0 & 0 & 0 & 0 & 0 & 0 & 0 & \Gamma & 0 \\ 
0 & 0 & 0 & 0 & 0 & 0 & 0 & 0 & \Gamma \\ 
\end{pmatrix} ,
\end{equation}




\noindent where in all cases there is a single non-zero free parameter, $\Gamma$, which has units of the inverse of distance, or equivalently energy.

To understand these $D$ matrices, it is useful to consider the resulting form of $\mathcal{D}[\rho]$. Ultimately $\mathcal{D}[\rho]$ is a $N \times N$ matrix\footnote{This is can be seen in \Cref{eqn:decoh_master}, where it is evident that $\mathcal{D}[\rho]$ has the same dimensions as $H$ and $\rho$.}, e.g.\@ $3 \times 3$ for a three neutrino system. It can be shown from \Cref{eqn:decoh_operator_sun} that $\Gamma_{3,8}$ determine the diagonal elements of $\mathcal{D}[\rho]$, whilst $\Gamma_{1,2,4,5,6,7}$ determine the off-diagonal elements. 

In the mass basis, standard neutrino oscillations are driven by a diagonal $H$ (resulting from non-zero mass splittings). These oscillatory terms appear as off-diagonal elements in the standard evolution term $i[H,\rho]$ in \Cref{eqn:decoh_master}, and thus oscillations cause time-dependence in the off-diagonal elements of $\rho$. These off-diagonal $\rho$ elements are damped to zero by non-zero $\Gamma_{1,2,4,5,6,7}$, damping the oscillations but preserving the diagonal $\rho$ elements that yield the PMNS matrix dependence of the large distance behaviour observed in the `phase perturbation' scenario. Non-zero $\Gamma_{3,8}$ instead produce damping in the diagonal (non-oscillatory) $\rho$ elements, which tend to the value $1/N$. In combination with the damped off-diagonal elements resulting from non-zero $\Gamma_{1,2,4,5,6,7}$, this produces the $1/N$ large distance behaviour observed for the `state selection` cases\footnote{The damping of non-oscillatory elements of $\rho$ is sometimes referred to as \textit{neutrino relaxation} in the literature~\cite{GUZZO2016408, Gomes:2020muc}.}. Finally, the addition of non-zero $\Gamma_0$ causes the diagonal $\rho$ elements to damp to 0 (instead of $1/N$). In this case, all $\rho$ elements tend to 0, resulting in the non-unitary `neutrino loss' scenario.

More generally, we note that the `state selection' case will represent state selection in \textit{any} basis, as for unitary mixing an equal population of mass states must correspond to equal populations of the mixed states.  This scenario is thus also sensitive to interactions selecting any new neutrino basis states (unrelated to the weak nuclear force) resulting from new physics. 

It is useful to note that the $\mathcal{D}[\rho]$ operator resulting from \Cref{eqn:randomized_phase_D} is:

\begin{equation}
\label{eqn:randomized_phase_Drho_NxN}
\mathcal{D}[\rho] = \begin{pmatrix}
0 & \Gamma \rho_{10} & \Gamma \rho_{20} \\ 
\Gamma \rho_{01} & 0 & \Gamma \rho_{21} \\ 
\Gamma \rho_{02} & \Gamma \rho_{12} & 0 \\ 
\end{pmatrix} ,
\end{equation}

\noindent which is a common form that has been explored in the literature~\cite{Farzan:2008zv,PhysRevD.96.093009,Coloma:2018idr}, and so these limits can be interpreted in terms of the $\nu$-VBH `phase perturbation' interactions considered here. More generally, the mapping of a diagonal $D$ matrix to $\mathcal{D}[\rho]$ when expressed as an $N \times N$ matrix is given by:

\begin{widetext}
\begin{equation}
\label{eqn:general_Drho_NxN}
\renewcommand\arraystretch{2}
\mathcal{D}[\rho] = \begin{pmatrix}
\Omega_0 + \Omega_3 + \Omega_8 & \Gamma_1 \operatorname{Re}\{\rho_{01}\} - i \Gamma_2 \operatorname{Im}\{\rho_{10}\} & \Gamma_4 \operatorname{Re}\{\rho_{02}\} - i \Gamma_5 \operatorname{Im}\{\rho_{20}\} \\ 
\Gamma_1 \operatorname{Re}\{\rho_{01}\} + i \Gamma_2 \operatorname{Im}\{\rho_{10}\} & \Omega_0 - \Omega_3 + \Omega_8 & \Gamma_6 \operatorname{Re}\{\rho_{12}\} - i \Gamma_7 \operatorname{Im}\{\rho_{21}\} \\ 
\Gamma_4 \operatorname{Re}\{\rho_{02}\} + i \Gamma_5 \operatorname{Im}\{\rho_{20}\} & \Gamma_6 \operatorname{Re}\{\rho_{12}\} + i \Gamma_7 \operatorname{Im}\{\rho_{21}\} & \Omega_0 - 2 \Omega_8 \\
\end{pmatrix} ,
\end{equation}
\end{widetext}

where the $\Omega_\mu$ terms are given by:

\begin{equation}
\begin{split}
\Omega_0 & = \frac{\Gamma_0}{3} \left ( \rho_{00} + \rho_{11} + \rho_{22} \right ), \\
\Omega_3 & = \frac{\Gamma_3}{2} \left ( \rho_{00} - \rho_{11} \right ), \\
\Omega_8 & = \frac{\Gamma_8}{6} \left ( \rho_{00} + \rho_{11} - 2 \rho_{22} \right ). \\
\end{split}
\end{equation}

This mapping\footnote{The specific case of \Cref{eqn:randomized_phase_Drho_NxN} results when $\Gamma_{1,2,4,5,6,7} = \Gamma$ and $\Gamma_{0,3,8} = 0$, in addition to the properties $\operatorname{Re}\{\rho_{ij}\} = \operatorname{Re}\{\rho_{ji}\}$ and $\operatorname{Im}\{\rho_{ij}\} = -\operatorname{Im}\{\rho_{ji}\}$.} is useful for comparing forms of $\mathcal{D}[\rho]$ expressed with and without the $SU(N)$ expansion described in Section \ref{sec:open_quantum_system_decoh}.

The $D$ matrices given by \Cref{eqn:randomized_flavor_D,eqn:randomized_phase_D,eqn:neutrino_loss_D} produce damping terms of the form $e^{-\Gamma L}$. To attribute physical meaning to the value of $\Gamma$, we define the coherence length, $L_{\rm{coh}}$, of the ensemble as the distance at which damping terms have reached $e^{-1}$, which implies:

\begin{equation}
\label{eqn:coherence_length}
L_{\rm{coh}} = \frac{1}{\Gamma}.
\end{equation}

Since the $\nu$-VBH interaction cases considered in this work produce a total loss of coherence after a single interaction (e.g.\@ the final state is independent of the initial state), $L_{\rm{coh}}$ is equal to the interaction mean free path, and experimental constraints on $\Gamma$ (and thus $L_{\rm{coh}}$) can therefore be directly interpreted as constraints on the mean free path of $\nu$-VBH interactions.

\begin{figure} 
    \centering
    \includegraphics[trim=0.5cm 0.0cm 0.1cm 0.0cm, clip=true, width=\linewidth]{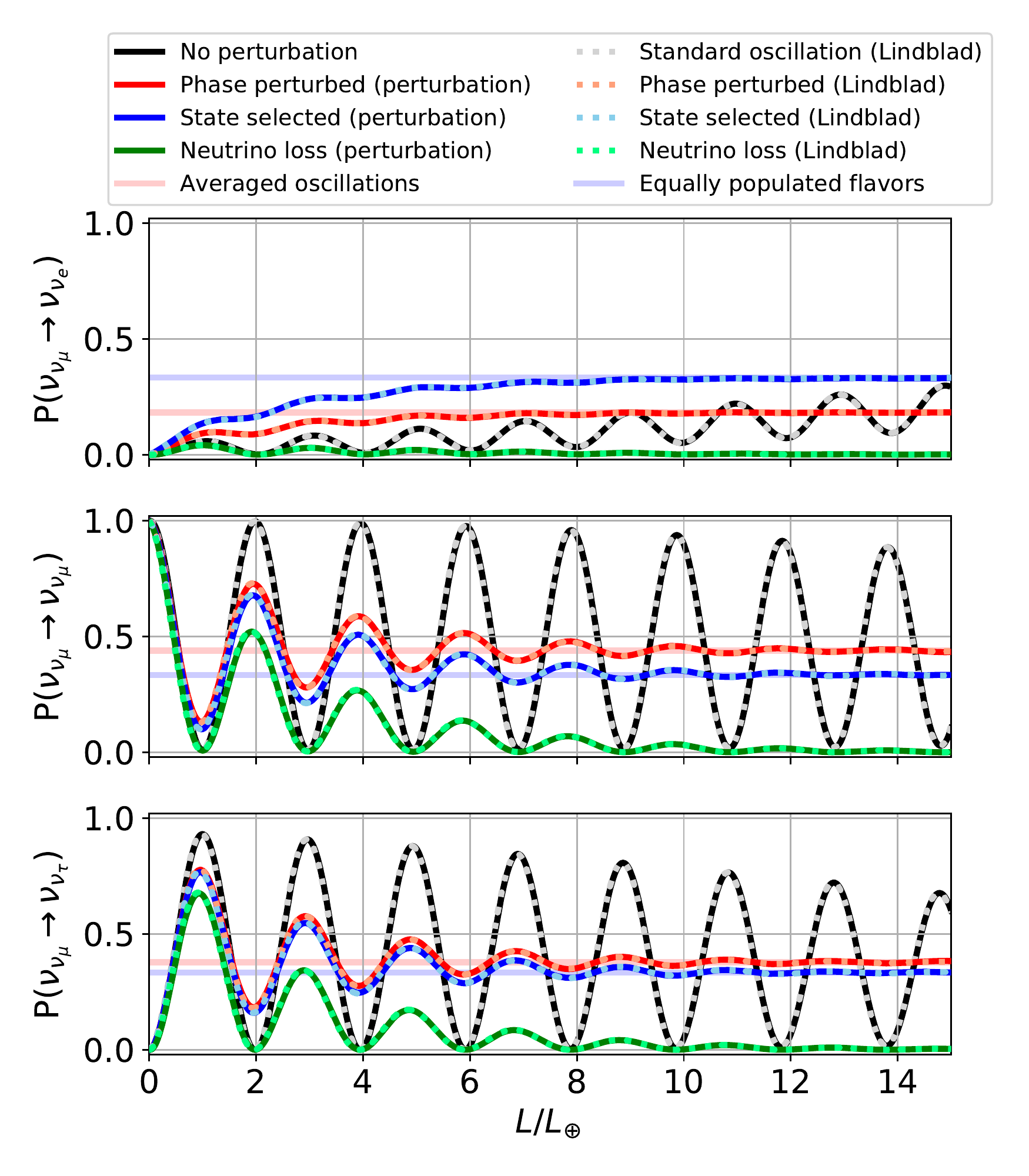}
    \caption{Oscillation probability resulting from $\nu$-VBH interactions, computed both by perturbing propagating neutrinos and using the Lindblad open quantum system formalism. A 3 neutrino system is shown in vacuum with the oscillation parameters in \Cref{table:3nu_params}. $L_{\oplus}$ is the diameter of Earth. Both the $\nu$-VBH interaction mean free path and $1/\Gamma$ are set to $3L_{\oplus}$.}
    \label{fig:perturbation_vs_lindblad}
\end{figure}

To verify the $D$ matrices in \Cref{eqn:randomized_flavor_D,eqn:randomized_phase_D,eqn:neutrino_loss_D} and also the assertion that $L_{\rm{coh}}$ can be interpreted as the $\nu$-VBH interaction mean free path, in \Cref{fig:perturbation_vs_lindblad} we show the oscillation probabilities computed using both the open quantum system formalism and by injecting perturbations into our neutrino propagation model (as described in Section \ref{sec:modelling_nuVBH_interactions}). A 3 neutrino system with realistic oscillation parameters is shown, with the injected coherence length shown being of relevance to quantum gravity searches with atmospheric neutrinos. We observe perfect agreement between the two approaches in all cases, and conclude that the open quantum system models presented in this section do indeed correctly represent the $\nu$-VBH interaction scenarios investigated, and can be used to experimentally search for quantum gravity. The open quantum system model is implemented in the \texttt{nuSQuIDS} software package~\cite{Delgado:2014kpa, nusquidsGIT}, and is solved numerically.

\section{\label{sec:planck_scale_decoh}Energy-dependence of decoherence from Planck scale physics}

The general open quantum system approach outlined in Section \ref{sec:open_quantum_system_decoh} does not implicitly consider the energy dependence of the physics producing the decoherence effects, i.e. the $\nu$-VBH interactions in this work. This can be introduced however by defining the energy-dependence of the free parameters in $D$. As previously stated, there is currently no generally accepted theory of quantum gravity, and so we instead take a phenomenological approach and introduce a general form for the energy-dependence of the $\Gamma$ parameter controlling the decoherence effects. A common approach in the literature has been to assume a power-law energy-dependence~\cite{PhysRevLett.85.1166,PhysRevD.76.033006,GUZZO2016408,Gomes:2020muc}:

\begin{equation}
\label{eqn:gamma_energy_dep}
\Gamma(E) = \Gamma(E_0) \left( \frac{E}{E_0} \right)^n ,
\end{equation}

\noindent where $E_0$ is a reference energy pivot and $n$ is the power-law index. Studies often test multiple cases for $n$, rather than assuming a specific model.

As an aside, an interesting observation is that the case of $n=-1$ coupled with the `neutrino loss' $D$ matrix shown in in \Cref{eqn:neutrino_loss_D} produces a signal that is phenomenologically identical to neutrino decay with invisible decay products (where the energy-dependence results from time dilation).

Noting that $\Gamma$ has units of energy, \Cref{eqn:gamma_energy_dep} can be re-written to express the $\Gamma$ parameters with respect to an arbitrary energy scale, $\Lambda$:

\begin{equation}
    \label{eqn:gamma_lambda}
    \Gamma(E) = \zeta \frac{E^n}{\Lambda^{n-1}} ,
\end{equation}

\noindent where $\zeta$ is a dimensionless constant, and is a free parameter characterising the strength of the decoherence effects.

When considering decoherence from quantum gravity, the energy scale of interest is the Planck mass, $\Lambda \sim M_{\rm{Planck}} \simeq 1.2 \times 10^{19}$\,GeV, and thus $\Gamma$ can be expressed relative to the Planck scale as~\cite{Anchordoqui:2005gj}:

\begin{equation}
    \label{eqn:gamma_planck}
    \Gamma(E) = \zeta_{\rm{Planck}} \frac{E^n}{M_{\rm{Planck}}^{n-1}} .
\end{equation}

Using \Cref{eqn:coherence_length}, \Cref{eqn:gamma_planck} can also be expressed as an energy-dependent coherence length relative to the Planck length, $L_{\rm{Planck}}$:

\begin{equation}
    \label{eqn:coherence_length_planck_scale}
    L_{\rm{coh}}(E) = \frac{ L_{\rm{Planck}} }{\zeta_{\rm{Planck}}} \left( \frac {M_{\rm{Planck}} }{ E } \right)^n .
\end{equation}

\Cref{eqn:coherence_length_planck_scale} yields physical insight into this energy-dependence parameterisation. From it, we see that a neutrino with $E = M_{\rm{Planck}}$ would have a coherence length of $\zeta^{-1}_{\rm{Planck}}$ Planck lengths, regardless of $n$.  $\zeta^{-1}_{\rm{Planck}}$ can thus be interpreted as the neutrino coherence length at the Planck scale, whilst the $\left( M_{\rm{Planck}} / E  \right)^n$ term encodes the suppression of the decoherence effects at neutrino energies below from the Planck scale. In general, theories of quantum gravity predict significant effects at the Planck scale with large suppression at lower energy scales, which can be represented using \Cref{eqn:gamma_planck,eqn:coherence_length_planck_scale} when $n>0$. As such only positive $n$ are considered for the remainder of this section.

\begin{figure}
    \centering
    \includegraphics[trim=0.0cm 0.2cm 0.cm 0.2cm, clip=true, width=\linewidth]{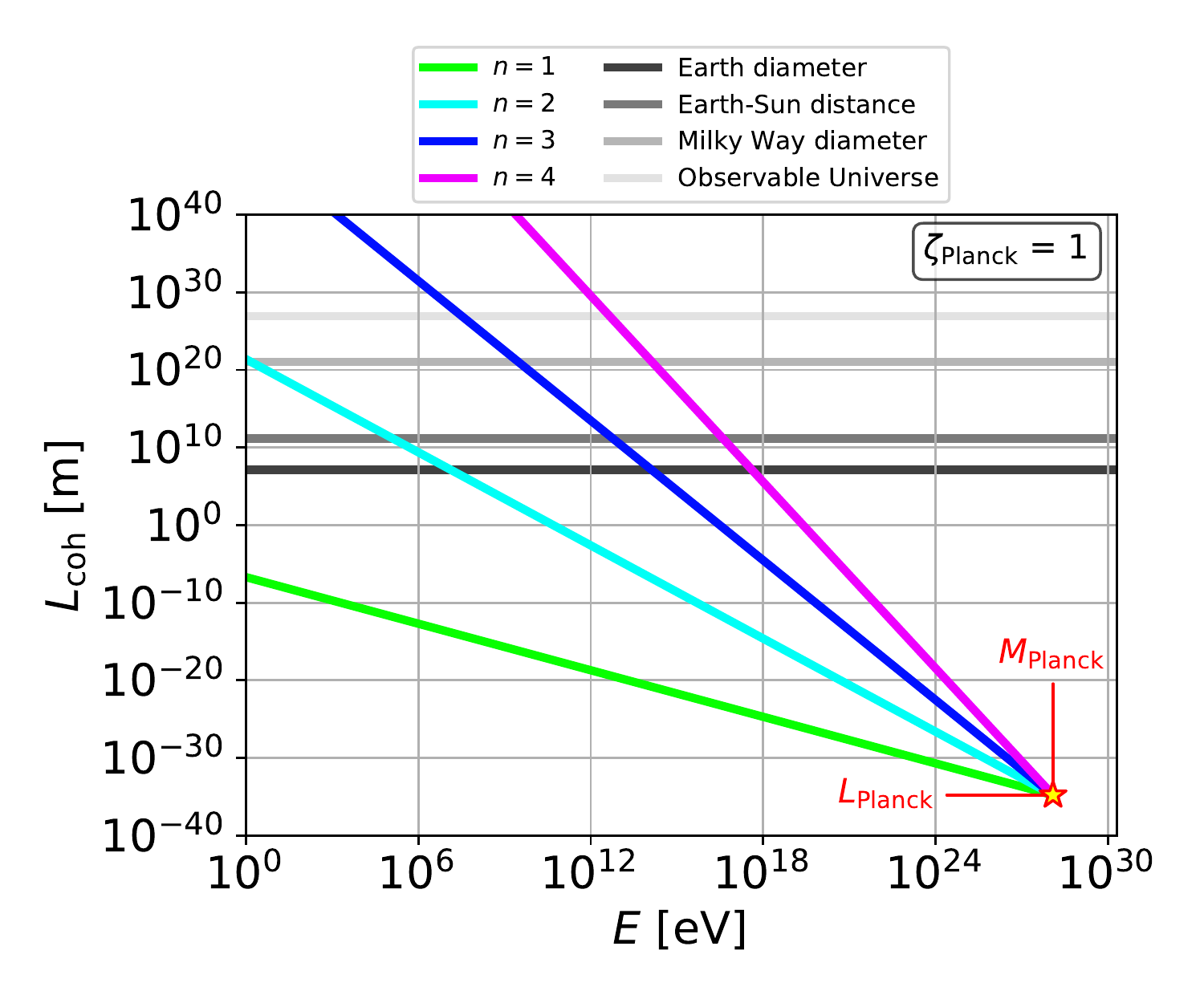}
    \caption{Neutrino coherence length versus neutrino energy resulting from a `natural' ($\zeta_{\rm{Planck}} = 1$) Planck scale source of decoherence, for a range of possible $n$ cases. Reference distance scales are shown for comparison.}
    \label{fig:planck_decoh_distance}
\end{figure}

A `natural' Planck scale theory is expected to have $\zeta_{\rm{Planck}} \sim \mathcal{O}(1)$~\cite{Anchordoqui:2005gj}. \Cref{fig:planck_decoh_distance} shows the coherence length as a function of neutrino energy predicted by  \Cref{eqn:coherence_length_planck_scale} under this assumption of naturalness for a range of $n$. For all $n$ tested, coherence length decreases with increasing neutrino energy as the suppression of Planck scale effects at low energies diminishes, and ultimately all cases converge at the Planck scale, where the coherence length $\sim L_{\rm{Planck}}$. Lower $n$ produces smaller coherence lengths (e.g.\@ stronger decoherence effects) at any given sub-Planck energy, as lower $n$ represents weaker suppression.

A number of reference distances are shown for comparison to the predicted natural coherence lengths in \Cref{fig:planck_decoh_distance}. We see that the weakest suppression, i.e.\@ $n=1$, case predicts microscopic coherence lengths for all energies probed by neutrino experiments, and is thus strongly excluded at this natural scale by the non-detection of neutrino decoherence in any experiment to date. Note that exclusion at the natural scale does not exclude the model outright, but  constrains it to $\zeta_{\rm{Planck}} < 1$ (by many orders of magnitude in this case). At the other extreme, if $n=4$ then neutrinos of up to $\sim$TeV energies have natural coherence lengths larger than the observable Universe, making decoherence effects essentially unobservable. However, significant decoherence would occur for the high-energy extragalactic neutrino flux observed by neutrino telescopes such as IceCube~\cite{Aartsen:2013jdh} that extends into the PeV and even EeV range (detectable by radio neutrino detectors and cosmic ray air shower detectors in the case of Earth skimming neutrinos).

The case of $n=2$ is of particular interest as this energy-dependence has been predicted by work probing quantum decoherence effects in string theory models, including for particles encountering black holes in four dimensions~\cite{Ellis:1996bz} and D-brane foam backgrounds~\cite{Ellis:1997jw,BENATTI199958}. We see from \Cref{fig:planck_decoh_distance} that a natural $n=2$ Planck scale theory predicts a neutrino coherence length smaller than the Earth's diameter for $E \gtrsim$10\, MeV, and $\sim$1\,km at 1\,GeV. Such strong decoherence effects have not been observed by long baseline accelerator and atmospheric neutrino experiments, constraining any such theory well below the natural scale ($\zeta_{\rm{Planck}} \ll 1$). 

We can comment on what a notional `natural' theory really represents in the case of the $\nu$-VBH interactions considered in this work. From \Cref{eqn:coherence_length_planck_scale}, the `natural' case of $\zeta_{\rm{Planck}} \sim 1$ implies  $L_{\rm{coh}} \sim L_{\rm{Planck}}$ for a neutrino with Planck scale energies, which according to the conclusions derived in Section \ref{sec:oqs_vs_perturbation} implies a $\nu$-VBH interaction occurs, on average, every Planck length travelled by a Planck scale neutrino. If $\nu$-VBH interactions are less frequent than this, it would imply $\zeta_{\rm{Planck}} < 1$ and thus weaker signals at the energies probed by neutrino experiments, potentially evading detection thus far.

Ultimately, $\zeta_{\rm{Planck}}$ is a free parameter that must be measured or constrained using experimental data. Experimental constraints on $\Gamma(E_0)$ from analyses using the energy-dependence parameterisation given by \Cref{eqn:gamma_energy_dep} can be converted to $\zeta_{\rm{Planck}}$ as follows:

\begin{equation}
    \label{eqn:gamma_planck_vs_pheno}
    \zeta_{\rm{Planck}} = \Gamma(E_0) \frac{M_{\rm{Planck}}^{n-1}}{E_0^n} .
\end{equation}

For example, the limit\footnote{Note that this result considers only 2 neutrino flavors.} of $\Gamma(E_0) < 0.9\times10^{-27}$\,GeV ($n=2$) derived using data from the SuperKamiokande experiment~\cite{PhysRevLett.85.1166} corresponds to $\zeta_{\rm{Planck}} < 1.1\times10^{-8}$. 


\section{\label{sec:astro_atmo_nu}Decoherence in astrophysical and atmospheric neutrinos}

Now that we have mathematical definitions for neutrino decoherence and other effects resulting from $\nu$-VBH interactions, including their energy-dependence, we can evaluate the resulting potential signals in neutrino detectors. Given that decoherence effects accumulate over distance (until coherence is fully lost) and that Planck scale physics is expected to be suppressed at energies below the Planck scale, decoherence effects from quantum gravity are expected to manifest most strongly in neutrinos with high energies and long propagation baselines.

The diffuse extragalactic high-energy neutrino flux discovered by the IceCube neutrino observatory~\cite{Aartsen:2013jdh} initially seems an ideal hunting ground for such physics. Neutrinos of up to $\sim$PeV energies have been observed, and evidence has been found of neutrinos travelling $\sim$Gpc distances~\cite{IceCube:2018cha,IceCube:2018dnn}. The very fact that neutrinos from such distances have been observed at all significantly constrains the `neutrino loss' scenario considered in this work, but quantitative statements are however difficult without a detailed knowledge of the nature and distribution of sources, not to mention the neutrino flux they produce.

\begin{figure}
    \centering
    \includegraphics[trim=0.0cm 0.1cm 0.cm 0.2cm, clip=true, width=\linewidth]{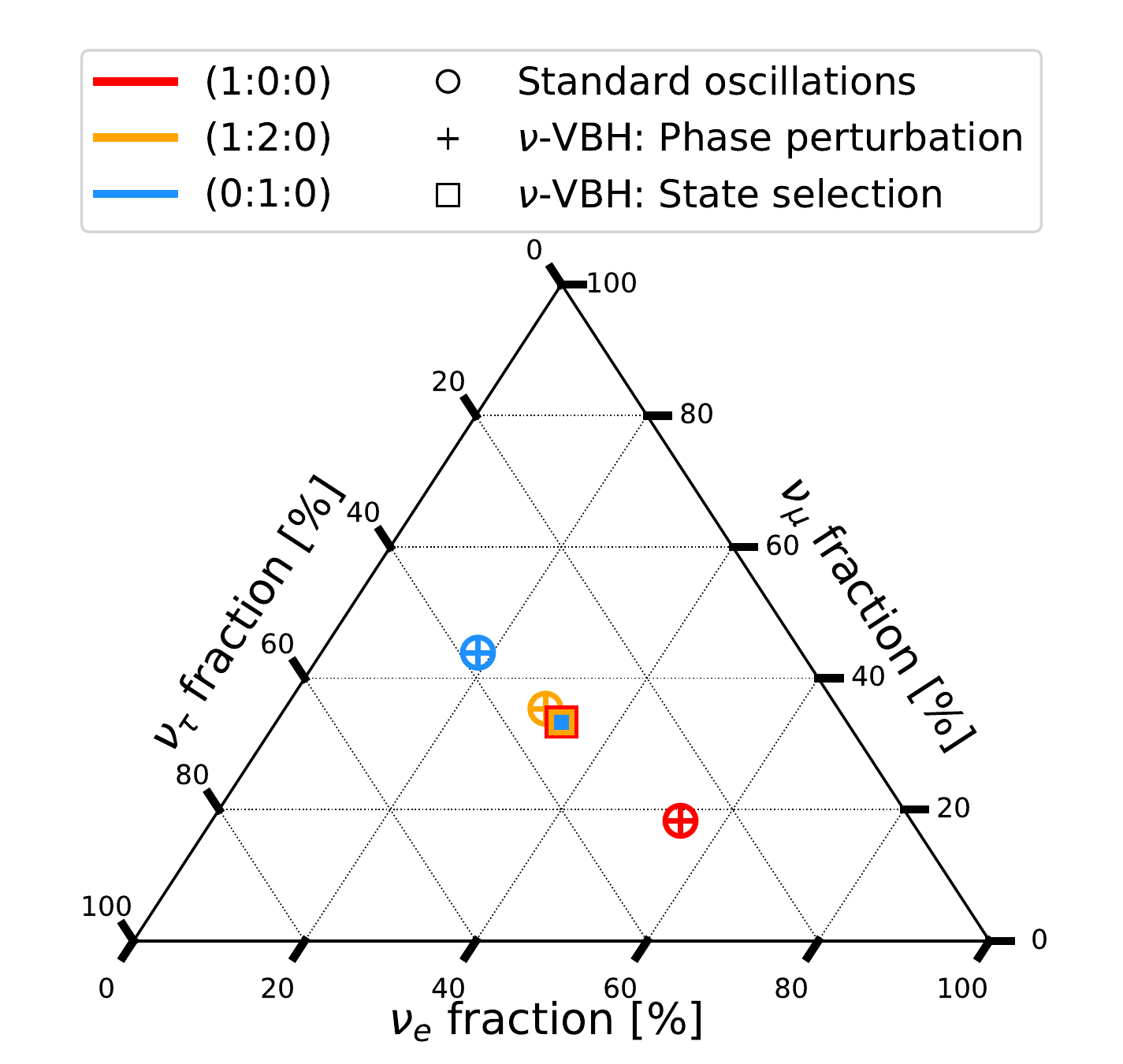}
    \caption{Astrophysical neutrino flavor triangle showing the ratios of each neutrino flavor expected at Earth from the diffuse astrophysical neutrino flux, for a range of possible initial source flux flavor ratios. Three different possible source flavor ratios are shown: $(\nu_{e}:\nu_{\mu}:\nu_{\tau})$ = $(1:0:0)$ (neutron decay), $(1:2:0)$ (pion decay), and $(0:1:0)$ (muon-damped pion decay). Both the standard oscillation expectation and $\nu$-VBH decoherence cases are shown. Oscillation parameters from \Cref{table:3nu_params} are used. }
    \label{fig:flavor_triangle}
\end{figure}

However, there is another fundamental limitation in observing $\nu$-VBH interactions from the diffuse astrophysicical neutrino flux. Due to the large and unknown travel distances, as well as broad energy distributions and finite detector resolution, the neutrinos are observed at Earth in an oscillation-averaged state~\cite{PhysRevD.98.123023}. This however is also precisely the long distance result of the `phase perturbation' $\nu$-VBH interactions described in this work. A fully decohered diffuse astrophysical neutrino flux is thus indistinguishable from the no-decoherence expectation. This is shown in \Cref{fig:flavor_triangle}, which shows the expected terrestrial neutrino flavor ratio (presented as a \textit{flavor triangle}) for a number of different source flux cases.

\Cref{fig:flavor_triangle} also demonstrates a major challenge in measuring the `mass/flavor state selection' $\nu$-VBH interactions described in this work with diffuse astrophysical neutrinos. The long range behaviour in this case is equally populated neutrino flavors, which produces a 1:1:1 flavor ratio at the Earth (assuming democratic flavor selection and full loss of coherence) regardless of initial flux. This is almost identical to the standard oscillation expectation for a pion decay source, and thus this case is also indistinguishable from the standard expectation with our present level of uncertainty as to the mechanisms producing the astrophysical neutrino flux. More differentiable signals could exist for coherence lengths approximately commensurate with the neutrino propagation distance, where coherence would not be completely lost at the Earth, or for undemocratic flavor scenarios. 

Neutrinos from identified astrophysical objects (\textit{`point sources'}) could in principal offer sensitive searches for neutrino decoherence, provided that they have a well understood initial neutrino flux, a well measured distance, and have a compact neutrino emission region. However, the identification of such sources at high energies, either galactic or extragalactic, is still in its infancy, with only a single source being identified thus far with $\mathcal{O}$(10) neutrinos associated with it~\cite{IceCube:2018cha}. Additionally, neutrinos identified as originating from distant point-like sources are typically $\nu_\mu$ events producing long muon tracks in detectors in charged-current interactions, as only these have sufficient direction resolution. Neutrino source identification is thus not sensitive at present to other neutrino flavours, unless observed via time coincidence with a short time-scale flaring event. Coupled with the large uncertainties in modelling the source neutrino flux (both steady-state and in transient flaring emission periods) and finite detector resolution, it is not yet possible to do robust quantitative searches for new physics based on the flavor composition of neutrino point source emission. Upper bounds on the mean free path of the `neutrino loss' scenario however can be derived from the fact that neutrinos are observed at all from a source of known distance~\cite{Kelly_2018}.

The Sun or galactic supernovae both represent other astrophysical neutrino source candidates, but have neutrino emission of $\mathcal{O}$(MeV), limiting sensitivity to Planck scale effects. Solar atmospheric neutrinos~\cite{Seckel:1990pc,Arg_elles_2017,PhysRevD.96.103006,Edsj__2017,Gutlein:2010ba} are higher energy, but again their study is in its infancy.

Closer to home, atmospheric neutrinos do offer a compelling source for neutrino decoherence studies. They offer a copious and relatively well understood flux of neutrinos reaching TeV energies, travel distances of up to $\sim$12700\,km (the Earth's diameter, $L_\oplus$) - more than an order of magnitude greater than any existing accelerator experiment - and high statistics samples can be collected by large underground detectors. The range of baselines and energies detected also allows the damping or energy-dependence of any detected signal to be explored. Whilst the distances involved are unlikely enough to produce detectable effects from e.g.\@ lightcone fluctuations due to the fluctuating space-time metric, if some fraction of the neutrinos detected encounter VBHs as they cross the Earth a measurable signal could be produced.

\begin{figure}
    \centering
    \includegraphics[trim=0.5cm 0.0cm 0.5cm 0.0cm, clip=true, width=\linewidth]{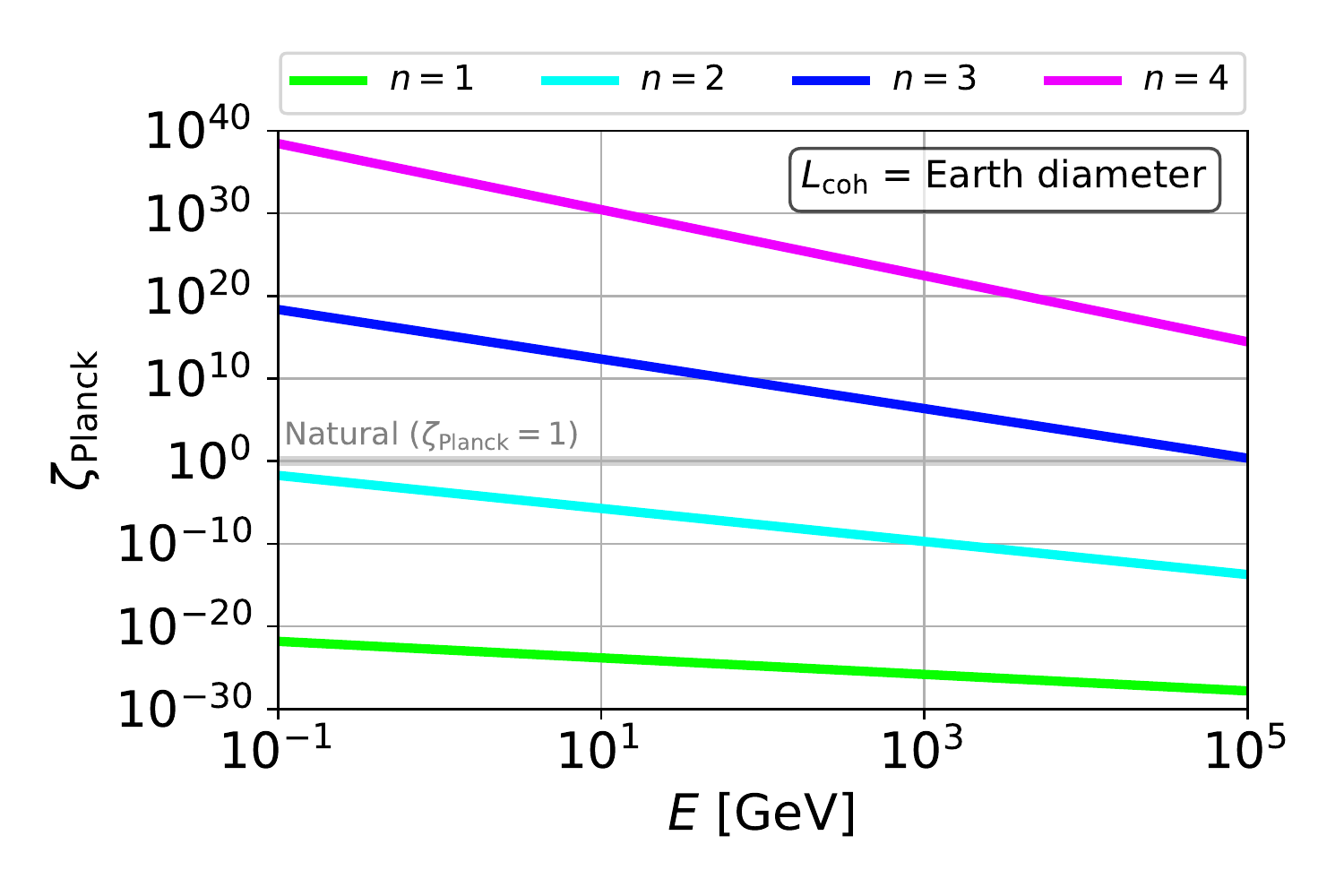}
    \caption{$\zeta_{\rm{Planck}}$ producing Earth diameter scale coherence lengths, shown for neutrinos energies relevant for atmospheric neutrinos and for a range of possible $n$ cases.}
    \label{fig:planck_atmo}
\end{figure}

\Cref{fig:planck_atmo} shows the $\zeta_{\rm{Planck}}$ value that produces decoherence with an Earth diameter scale neutrino coherence length, for a range of neutrino energies relevant for atmospheric neutrino experiments up to $\sim$100 TeV where the astrophysical flux starts to dominate\footnote{We note that inverting the self-veto techniques~\cite{PhysRevD.90.023009,Arguelles_2018} used to remove atmospheric neutrinos from astrophysical neutrino searches in neutrino telescopes could provide an enriched high-energy atmospheric neutrino sample above 100\,TeV for decoherence studies.}. We see that atmospheric neutrinos are sensitive to Planck scale physics many orders of magnitude below the natural expectation for the $n=1,2$ energy-dependence cases, and have limited sensitivity to $n=3$ too at the highest energies. 

\begin{figure}
    \centering
    \includegraphics[trim=0.0cm 0.5cm 0.cm 0.2cm, clip=true, width=\linewidth]{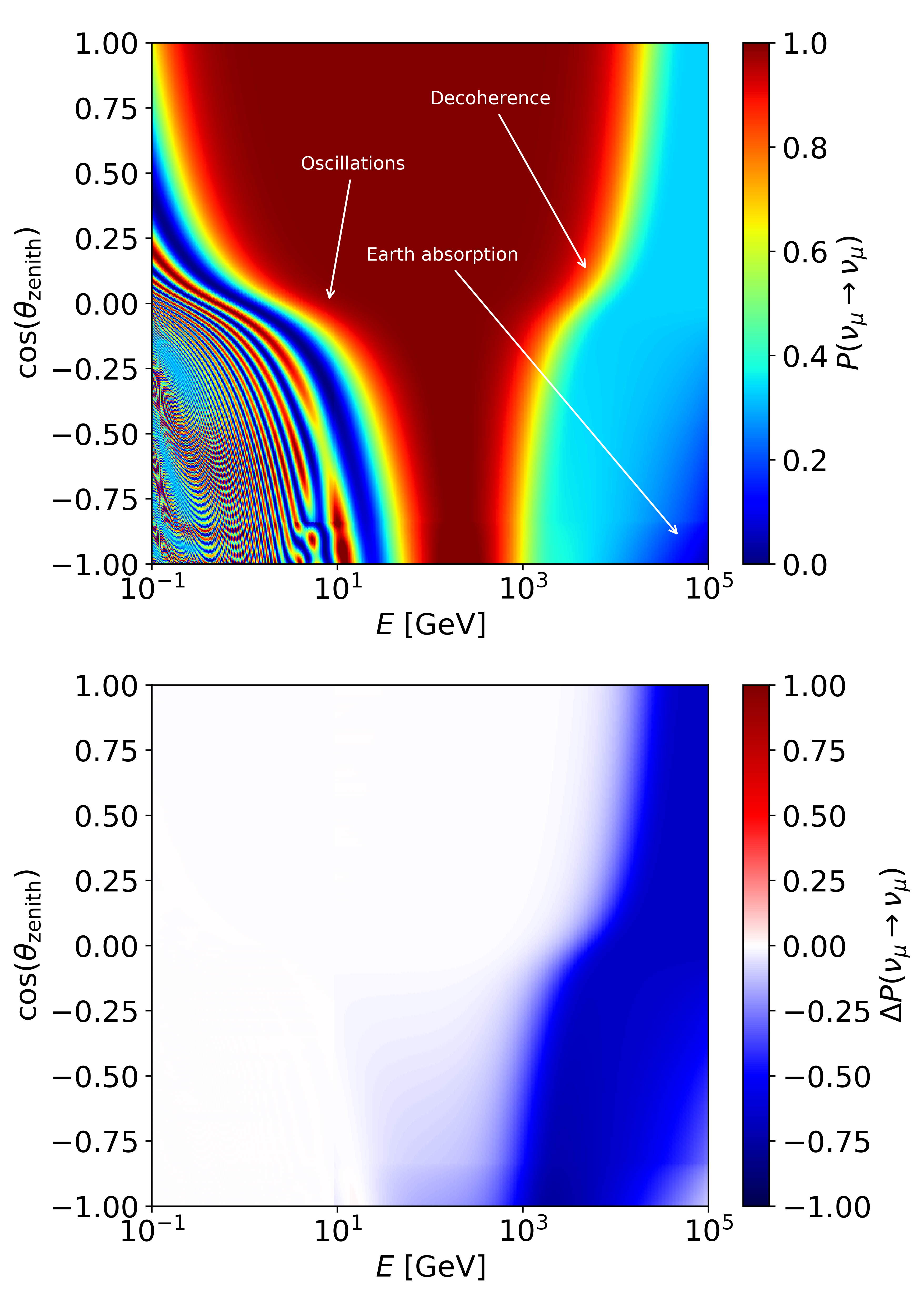}
    \caption{(Upper) Atmospheric $\nu_\mu$ survival probability in the presence of `flavor randomizing' $\nu$-VBH interactions with $L_{\rm{coh}}(\rm{1\:TeV}) = L_\oplus$ and $n=2$. Oscillation parameters from \Cref{table:3nu_params} are used. (Lower) The difference in oscillation probability w.r.t.\@ the standard (no decoherence case), $\Delta P = P_{\rm{decoherence}} - P_{\rm{standard}}$. }
    \label{fig:oscillogram_randomize_flavor_n2}
\end{figure}

The atmospheric neutrino oscillation probability in the presence of `state selection' $\nu$-VBH interactions is shown in \Cref{fig:oscillogram_randomize_flavor_n2} for the well-motivated $n=2$ case. The dominant $\nu_\mu$ disappearance channel is shown, as a function of the neutrino energy and (cosine of the) zenith angle; a proxy for atmospheric neutrino baseline. The oscillation parameters used throughout this section are given in \Cref{table:3nu_params}. Conventional matter effects are included as a potential in the Hamiltonian, with the matter density defined according to the Preliminary Reference Earth Model (PREM)~\cite{Dziewonski:1981xy}. The neutrino evolution is solved in the mass basis (using \texttt{nuSQuIDS}), and thus is not subject to issues introduced by approximations employed in other works (see \cite{Carpio:2017nui} for details). 

The $n>0$ cases motivated by Planck scale physics produce the strongest signals at higher energies, as can be seen for the $n=2$ cases in \Cref{fig:oscillogram_randomize_flavor_n2}. In this example, $\zeta_{\rm{Planck}}$ is chosen such that a 1\,TeV neutrino has a coherence length of one Earth diameter, producing a strong signal in a neutrino telescope such as IceCube and no significant signal in lower energy atmospheric or long baseline accelerator experiments. Full decoherence -- in this case equally populated flavors, i.e.\@ $P(\nu_\mu \rightarrow \nu_{e,\mu,\tau})=\frac{1}{3}$ -- results at high energies, even for down-going neutrinos ($\cos(\theta_{\rm zenith}) = 1$) travelling only $\mathcal{O}$(10\,km). In IceCube this would result in a large fraction of the dominant $\nu_\mu$ flux~\cite{Fedynitch:2017M4} at $>$1 TeV energies (observed as long muon tracks) being instead detected as $\nu_{e,\tau}$ events (observed as more spherical showers of light referred to as `cascades'), producing a clearly measurable signal given the high statistics neutrino samples IceCube has accumulated; $\mathcal{O}$($10^5$) $\nu_\mu$ at $\mathcal{O}$(TeV)~\cite{PhysRevLett.125.141801}. IceCube would thus be sensitive to decoherence signals even significantly weaker than this example. Note however that at the highest energies and longest distances, Earth absorption becomes significant, reducing the sensitivity to decoherence in precisely the region at which the signal would be strongest.

\begin{table}
 \begin{tabular}{|c| c|} 
 \hline
 Parameter & Value \\
 \hline
 \hline
 $\Delta m^2_{21}$ & $7.39 \times 10^{-5}$ $\rm{eV}^2$ \\ \hline
 $\Delta m^2_{31}$ & $2.528 \times 10^{-3}$ $\rm{eV}^2$ \\ \hline
 $\theta_{12}$ & 33.82$\degree$ \\ \hline
 $\theta_{13}$ & 8.60$\degree$ \\ \hline
 $\theta_{23}$ & 48.6$\degree$ \\ \hline
 $\delta_{CP}$ & 221$\degree$ \\ \hline
\end{tabular}
\caption{Neutrino oscillation parameters used for evaluating atmospheric neutrino oscillations, taken from NuFit 4.1 global fit results (normal mass ordering, SuperKamiokande data included)~\cite{NuFit41}.}
\label{table:3nu_params}
\end{table}

\begin{figure}
    \centering
        \centering
        \includegraphics[trim=0.0cm 12.7cm 0.cm 0.2cm, clip=true, width=\linewidth]{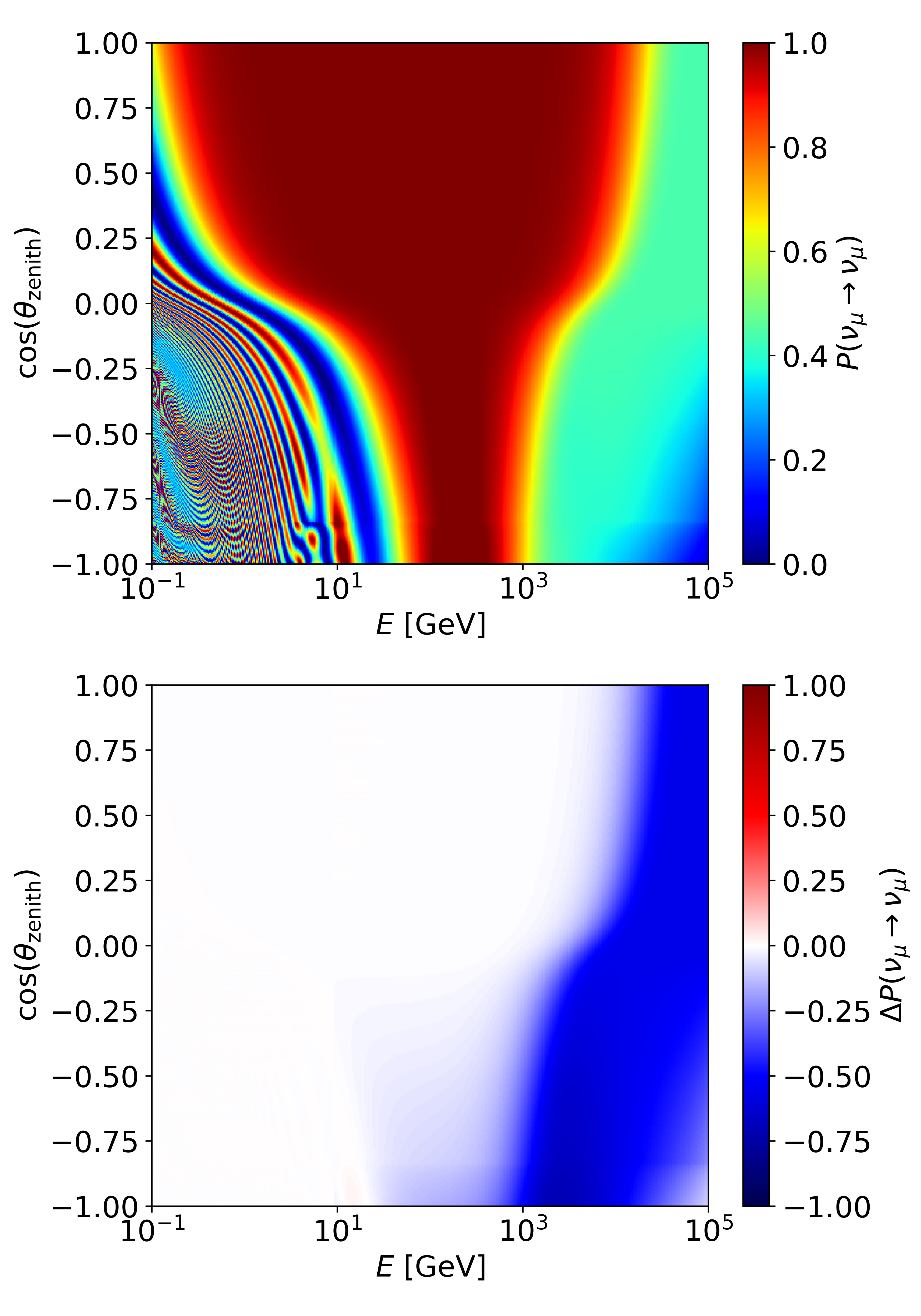}
        \centering
        \includegraphics[trim=0.0cm 13.0cm 0.cm 0.2cm, clip=true, width=\linewidth]{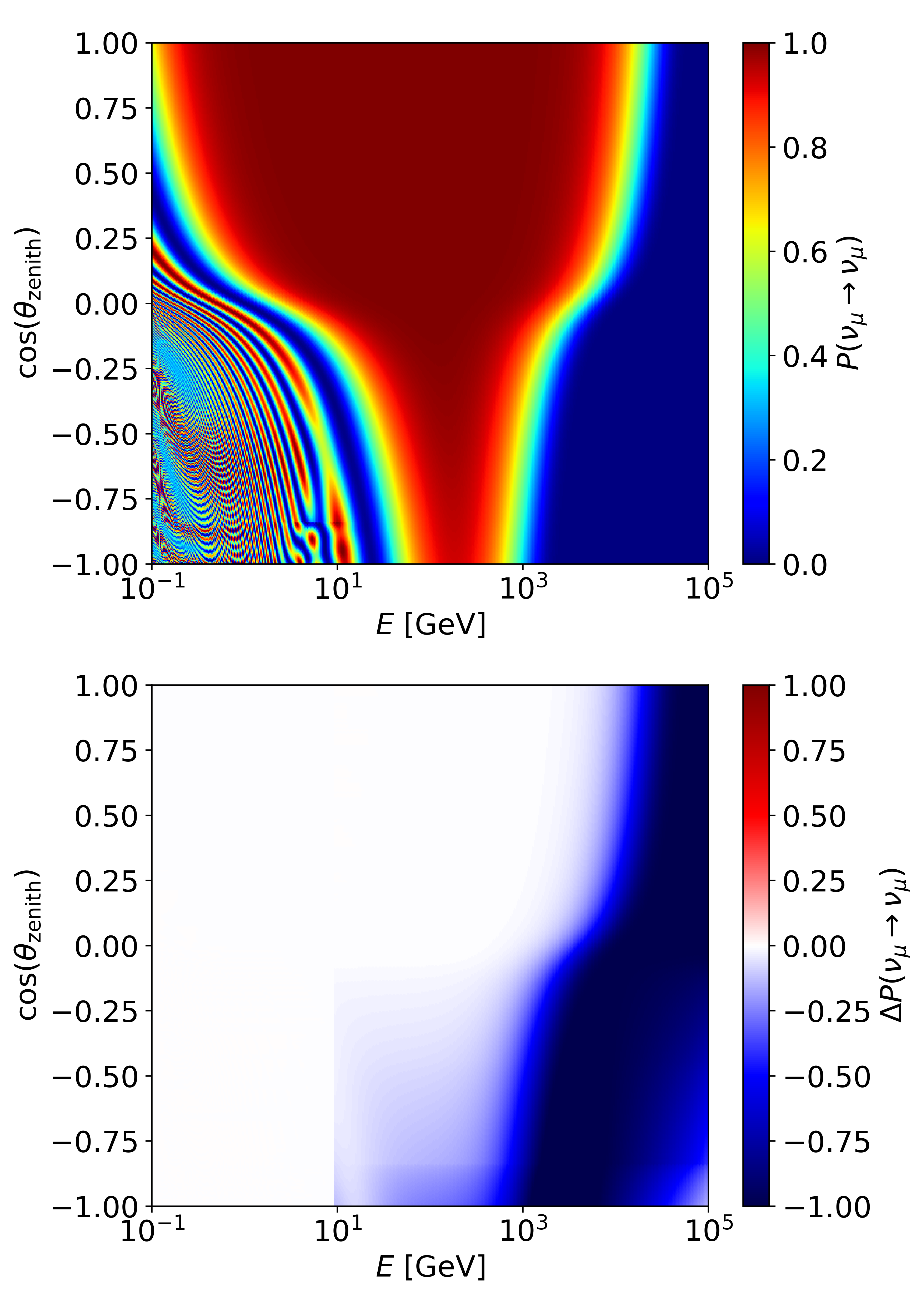}
    \caption{Atmospheric $\nu_\mu$ survival probability in the presence of `phase perturbation' (upper) and `neutrino loss' (lower) $\nu$-VBH interactions with $L_{\rm{coh}}(\rm{1\:TeV}) = L_\oplus$ and $n=2$ (same parameters as for \Cref{fig:oscillogram_randomize_flavor_n2}).}
    \label{fig:oscillogram_randomize_phase_neutrino_loss_n2}
\end{figure}

For comparison, the `phase perturbation' and `neutrino loss' $\nu$-VBH interaction cases for atmospheric neutrinos are shown in \Cref{fig:oscillogram_randomize_phase_neutrino_loss_n2}. As expected, similar trends are seen to the `mass/flavor state selection' case but with differing $P(\nu_\mu \rightarrow \nu_{\mu})$ once full decoherence is reached, as was seen in \Cref{fig:perturbation_vs_lindblad} in 1-dimension. Distinguishing between the different $\nu$-VBH cases may be difficult, as they produce similar signatures to each other, although with different apparent strengths. In particular the `phase perturbation' and `state selection` cases are similar due to maximal mixing in the atmospheric sector.

\begin{figure}
    \centering
    \includegraphics[trim=0.0cm 0.5cm 0.cm 0.2cm, clip=true, width=\linewidth]{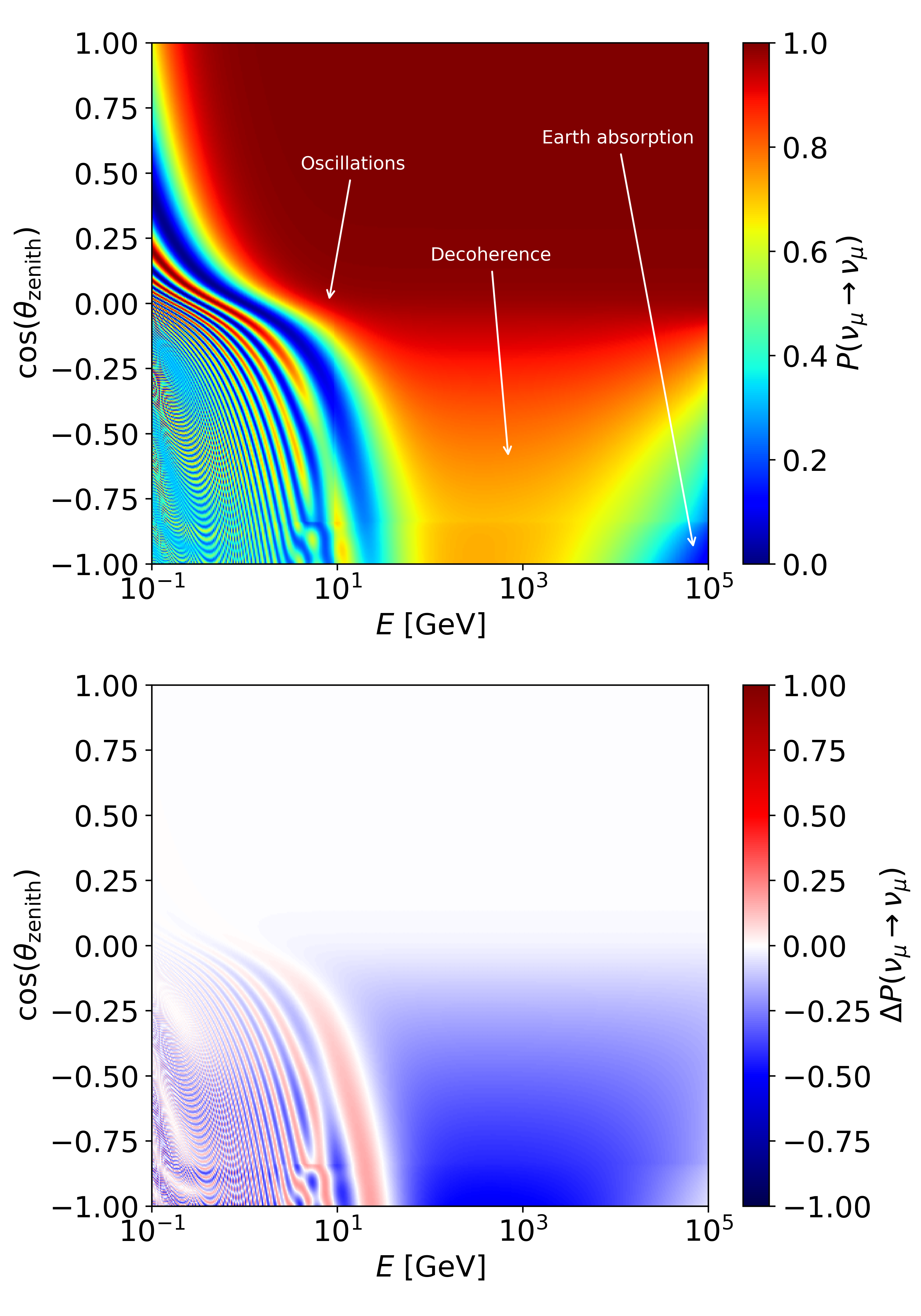}
    \caption{Plot analogous to \Cref{fig:oscillogram_randomize_flavor_n2} but with $n=0$ (energy-independent decoherence) and $L_{\rm{coh}}(\forall E) = L_\oplus$. }
    \label{fig:oscillogram_randomize_flavor_n0}
\end{figure}

Varying $n$ changes the energy at which the decoherence effects become significant. For comparison, in \Cref{fig:oscillogram_randomize_flavor_n0} we also show the decoherence signal resulting from energy-independent decoherence. Here the effects present most strongly at long baselines but across all energies, resulting in weaker oscillations in the standard oscillation region but introducing flavor transitions at higher energies where there are none in the standard picture. 

\begin{figure}
    \centering
    \includegraphics[trim=0.5cm 0.0cm 0.2cm 0.0cm, clip=true, width=\linewidth]{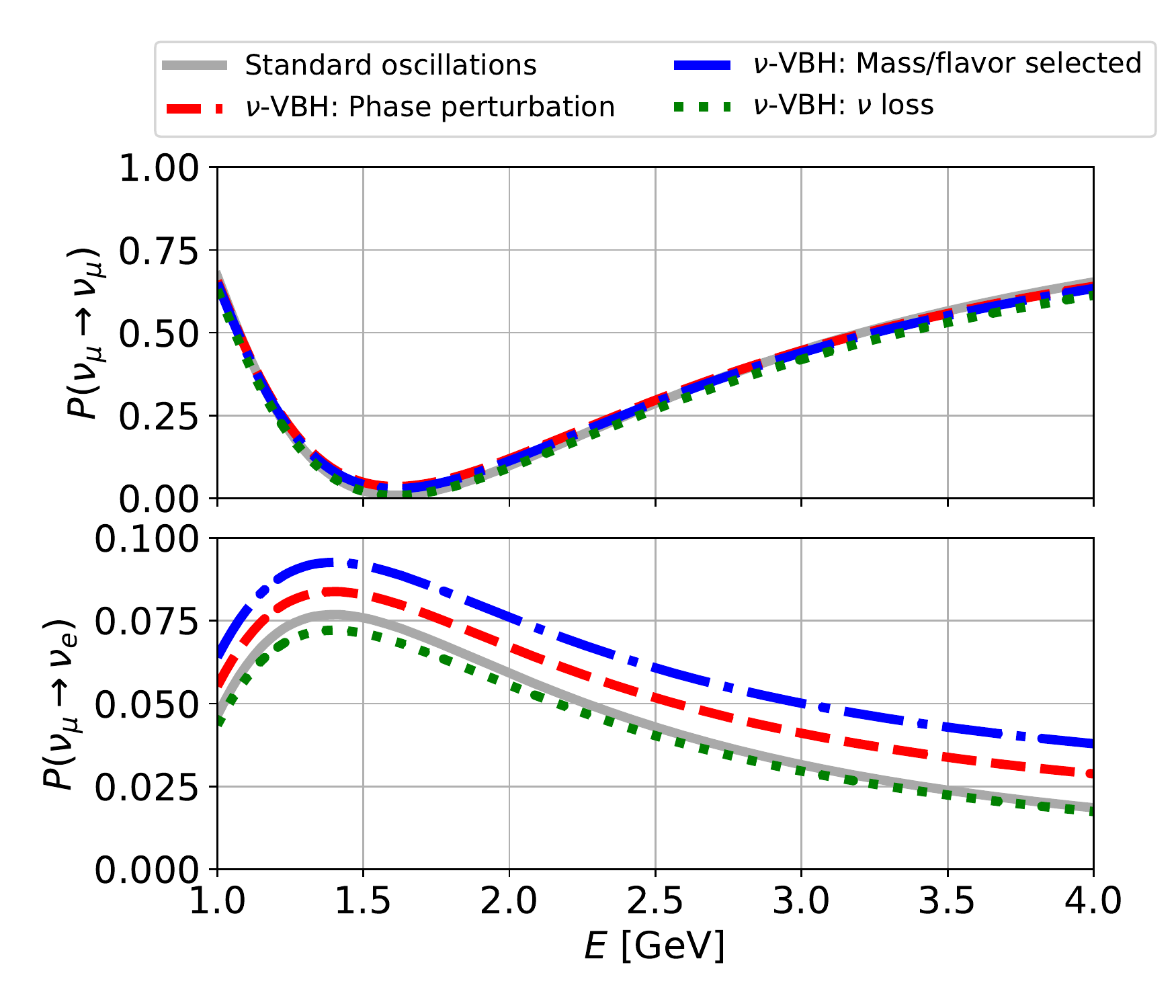}
    \caption{$\nu_\mu$ survival (upper) and $\nu_e$ appearance (lower) probability versus neutrino energy at the NO$\nu$A experiment, shown for various $\nu$-VBH interactions decoherence cases with $L_{\rm{coh}} = L_{\oplus}$ and $n=0$ (energy-independent). Parameters from \Cref{table:3nu_params} are used, and the experiment's baseline is 810\,km. Constant matter density of 2.84 $\rm{g/cm^3}$ is assumed~\cite{PhysRevLett.118.231801}. }
    \label{fig:lbl_atmo_decoh}
\end{figure}

This energy-independent case could in principle also produce measurable signals in long baseline accelerators (which operate at lower energies than e.g. IceCube), although the short baselines of these experiments compared to atmospheric neutrino experiments will result in significantly weaker effects. \Cref{fig:lbl_atmo_decoh} shows the oscillation probabilities for the NO$\nu$A experiment~\cite{Ayres:2007tu} corresponding to the atmospheric case shown in \Cref{fig:oscillogram_randomize_flavor_n0}, where the difference in oscillation probability with respect to standard oscillations being more than an order of magnitude smaller. The relative effect on the sub-leading $\nu_e$ appearance channel is of approximately the same scale as variations due to the value of $\delta_{CP}$ or uncertainty in $\theta_{13}$ however~\cite{T2K_CPPhase_Nature}. These energy-independent decoherence scenarios are not well motivated for quantum gravity searches however, and atmospheric neutrinos remain the most promising search arena.

\section{\label{sec:conclusions}Summary and conclusions}

In this work we have seen that perturbations to propagating neutrino states can cause neutrino decoherence and other modifications to neutrino flavor transitions, with stochasticity of the perturbation being a key ingredient for decoherence. Quantum gravity is postulated to provide a fluctuating environment that could induce such effects in neutrino propagation, and we have evaluated the impact of a range of heuristic $\nu$-VBH interaction scenarios, showing varied and potentially detectable and differentiable modifications to neutrino oscillation probability.

We have shown how the $\nu$-VBH interaction scenarios considered can be represented in the formalism of open quantum systems, allowing experimental decoherence constraints to be directly interpreted in terms of the mean free path of $\nu$-VBH interactions. In turn such results can constrain the VBH number density and the $\nu$-VBH interaction cross section. The heuristic interaction cases considered are relatively simple, and can potentially represent other new physics affects producing perturbations to propagating neutrino phases and other properties. Tantalisingly, we see that depending on the energy-dependence of the suppression of Planck scale physics at the energies our experiments operate at, sensitivity to Planck scale theories well below the `natural' scale can be achieved with current experiments, motivating further experimental searches.

Finally we have shown that high-energy diffuse astrophysical neutrinos are (counter-intuitively) of somewhat limited use in the search for $\nu$-VBH interactions via flavor transitions due to degeneracies with the standard flux expectation at the Earth, but that atmospheric neutrinos do however still offer power to probe decoherence signals many orders of magnitude below the natural Planck scale for cases motivated by string theory. With a new generation of atmospheric neutrino experiments currently under construction~\cite{Ishihara:2019aao, Adrian-Martinez:2016fdl, Abe:2011ts}, the prospects for Planck scale experimental physics are looking bright.


\section*{Acknowledgements}

\noindent The authors thank Jason Koskinen, Markus Ahlers, Carlos Arg\"{u}elles and Ben Jones for feedback on the paper draft, and these same people plus Subir Sarkar, Jo\~{a}o Coelho, Mauricio Bustamante, Shashank Shalgar, Mohamed Rameez, Peter Denton and Pilar Coloma for valuable conversations. This work is supported by VILLUM FONDEN (project no. 13161).



\nocite{*}

\bibliography{paper}

\begin{thebibliography}{77}%
\makeatletter
\providecommand \@ifxundefined [1]{%
 \@ifx{#1\undefined}
}%
\providecommand \@ifnum [1]{%
 \ifnum #1\expandafter \@firstoftwo
 \else \expandafter \@secondoftwo
 \fi
}%
\providecommand \@ifx [1]{%
 \ifx #1\expandafter \@firstoftwo
 \else \expandafter \@secondoftwo
 \fi
}%
\providecommand \natexlab [1]{#1}%
\providecommand \enquote  [1]{``#1''}%
\providecommand \bibnamefont  [1]{#1}%
\providecommand \bibfnamefont [1]{#1}%
\providecommand \citenamefont [1]{#1}%
\providecommand \href@noop [0]{\@secondoftwo}%
\providecommand \href [0]{\begingroup \@sanitize@url \@href}%
\providecommand \@href[1]{\@@startlink{#1}\@@href}%
\providecommand \@@href[1]{\endgroup#1\@@endlink}%
\providecommand \@sanitize@url [0]{\catcode `\\12\catcode `\$12\catcode
  `\&12\catcode `\#12\catcode `\^12\catcode `\_12\catcode `\%12\relax}%
\providecommand \@@startlink[1]{}%
\providecommand \@@endlink[0]{}%
\providecommand \url  [0]{\begingroup\@sanitize@url \@url }%
\providecommand \@url [1]{\endgroup\@href {#1}{\urlprefix }}%
\providecommand \urlprefix  [0]{URL }%
\providecommand \Eprint [0]{\href }%
\providecommand \doibase [0]{https://doi.org/}%
\providecommand \selectlanguage [0]{\@gobble}%
\providecommand \bibinfo  [0]{\@secondoftwo}%
\providecommand \bibfield  [0]{\@secondoftwo}%
\providecommand \translation [1]{[#1]}%
\providecommand \BibitemOpen [0]{}%
\providecommand \bibitemStop [0]{}%
\providecommand \bibitemNoStop [0]{.\EOS\space}%
\providecommand \EOS [0]{\spacefactor3000\relax}%
\providecommand \BibitemShut  [1]{\csname bibitem#1\endcsname}%
\let\auto@bib@innerbib\@empty
\bibitem [{\citenamefont {Fukuda}\ \emph {et~al.}(1998)\citenamefont {Fukuda}
  \emph {et~al.}}]{Fukuda:1998mi}%
  \BibitemOpen
  \bibfield  {author} {\bibinfo {author} {\bibfnamefont {Y.}~\bibnamefont
  {Fukuda}} \emph {et~al.} (\bibinfo {collaboration} {Super-Kamiokande}),\
  }\bibfield  {title} {\bibinfo {title} {Evidence for oscillation of
  atmospheric neutrinos},\ }\href@noop {} {\bibfield  {journal} {\bibinfo
  {journal} {Phys. Rev. Lett.}\ }\textbf {\bibinfo {volume} {81}},\ \bibinfo
  {pages} {1562} (\bibinfo {year} {1998})},\ \Eprint
  {https://arxiv.org/abs/hep-ex/9807003} {arXiv:hep-ex/9807003} \BibitemShut
  {NoStop}%
\bibitem [{\citenamefont {Ahmad}\ \emph {et~al.}(2001)\citenamefont {Ahmad}
  \emph {et~al.}}]{Ahmad:2001an}%
  \BibitemOpen
  \bibfield  {author} {\bibinfo {author} {\bibfnamefont {Q.~R.}\ \bibnamefont
  {Ahmad}} \emph {et~al.} (\bibinfo {collaboration} {SNO}),\ }\bibfield
  {title} {\bibinfo {title} {Measurement of the charged current interactions
  produced by b-8 solar neutrinos at the sudbury neutrino observatory},\
  }\href@noop {} {\bibfield  {journal} {\bibinfo  {journal} {Phys. Rev. Lett.}\
  }\textbf {\bibinfo {volume} {87}},\ \bibinfo {pages} {071301} (\bibinfo
  {year} {2001})},\ \Eprint {https://arxiv.org/abs/nucl-ex/0106015}
  {nucl-ex/0106015} \BibitemShut {NoStop}%
\bibitem [{\citenamefont {Ahmad}\ \emph {et~al.}(2002)\citenamefont {Ahmad}
  \emph {et~al.}}]{Ahmad:2002jz}%
  \BibitemOpen
  \bibfield  {author} {\bibinfo {author} {\bibfnamefont {Q.~R.}\ \bibnamefont
  {Ahmad}} \emph {et~al.} (\bibinfo {collaboration} {SNO}),\ }\bibfield
  {title} {\bibinfo {title} {Direct evidence for neutrino flavor transformation
  from neutral-current interactions in the sudbury neutrino observatory},\
  }\href@noop {} {\bibfield  {journal} {\bibinfo  {journal} {Phys. Rev. Lett.}\
  }\textbf {\bibinfo {volume} {89}},\ \bibinfo {pages} {011301} (\bibinfo
  {year} {2002})},\ \Eprint {https://arxiv.org/abs/nucl-ex/0204008}
  {nucl-ex/0204008} \BibitemShut {NoStop}%
\bibitem [{\citenamefont {Nussinov}(1976)}]{NUSSINOV1976201}%
  \BibitemOpen
  \bibfield  {author} {\bibinfo {author} {\bibfnamefont {S.}~\bibnamefont
  {Nussinov}},\ }\bibfield  {title} {\bibinfo {title} {Solar neutrinos and
  neutrino mixing},\ }\href
  {https://doi.org/https://doi.org/10.1016/0370-2693(76)90648-1} {\bibfield
  {journal} {\bibinfo  {journal} {Physics Letters B}\ }\textbf {\bibinfo
  {volume} {63}},\ \bibinfo {pages} {201 } (\bibinfo {year}
  {1976})}\BibitemShut {NoStop}%
\bibitem [{\citenamefont {Wolfenstein}(1978)}]{Wolfenstein:1977ue}%
  \BibitemOpen
  \bibfield  {author} {\bibinfo {author} {\bibfnamefont {L.}~\bibnamefont
  {Wolfenstein}},\ }\bibfield  {title} {\bibinfo {title} {{Neutrino
  Oscillations in Matter}},\ }\href {https://doi.org/10.1103/PhysRevD.17.2369}
  {\bibfield  {journal} {\bibinfo  {journal} {Phys. Rev.}\ }\textbf {\bibinfo
  {volume} {D17}},\ \bibinfo {pages} {2369} (\bibinfo {year} {1978})},\
  \bibinfo {note} {[,294(1977)]}\BibitemShut {NoStop}%
\bibitem [{\citenamefont {Mikheev}\ and\ \citenamefont
  {Smirnov}(1986)}]{Mikheev:1986wj}%
  \BibitemOpen
  \bibfield  {author} {\bibinfo {author} {\bibfnamefont {S.~P.}\ \bibnamefont
  {Mikheev}}\ and\ \bibinfo {author} {\bibfnamefont {A.~{\relax Yu}.}\
  \bibnamefont {Smirnov}},\ }\bibfield  {title} {\bibinfo {title} {{Resonant
  amplification of neutrino oscillations in matter and solar neutrino
  spectroscopy}},\ }\href {https://doi.org/10.1007/BF02508049} {\bibfield
  {journal} {\bibinfo  {journal} {Nuovo Cim.}\ }\textbf {\bibinfo {volume}
  {C9}},\ \bibinfo {pages} {17} (\bibinfo {year} {1986})}\BibitemShut {NoStop}%
\bibitem [{\citenamefont {Chizhov}\ \emph {et~al.}(1998)\citenamefont
  {Chizhov}, \citenamefont {Maris},\ and\ \citenamefont
  {Petcov}}]{Chizhov:1998ug}%
  \BibitemOpen
  \bibfield  {author} {\bibinfo {author} {\bibfnamefont {M.}~\bibnamefont
  {Chizhov}}, \bibinfo {author} {\bibfnamefont {M.}~\bibnamefont {Maris}},\
  and\ \bibinfo {author} {\bibfnamefont {S.}~\bibnamefont {Petcov}},\
  }\bibfield  {title} {\bibinfo {title} {{On the oscillation length resonance
  in the transitions of solar and atmospheric neutrinos crossing the earth
  core}},\ }\href@noop {} {\  (\bibinfo {year} {1998})},\ \Eprint
  {https://arxiv.org/abs/hep-ph/9810501} {arXiv:hep-ph/9810501} \BibitemShut
  {NoStop}%
\bibitem [{\citenamefont {Akhmedov}(1999)}]{Akhmedov:1998ui}%
  \BibitemOpen
  \bibfield  {author} {\bibinfo {author} {\bibfnamefont {E.~K.}\ \bibnamefont
  {Akhmedov}},\ }\bibfield  {title} {\bibinfo {title} {{Parametric resonance of
  neutrino oscillations and passage of solar and atmospheric neutrinos through
  the earth}},\ }\href {https://doi.org/10.1016/S0550-3213(98)00723-8}
  {\bibfield  {journal} {\bibinfo  {journal} {Nucl. Phys.}\ }\textbf {\bibinfo
  {volume} {B538}},\ \bibinfo {pages} {25} (\bibinfo {year} {1999})},\ \Eprint
  {https://arxiv.org/abs/hep-ph/9805272} {arXiv:hep-ph/9805272 [hep-ph]}
  \BibitemShut {NoStop}%
\bibitem [{\citenamefont {Hawking}(1982)}]{Hawking}%
  \BibitemOpen
  \bibfield  {author} {\bibinfo {author} {\bibfnamefont {S.~W.}\ \bibnamefont
  {Hawking}},\ }\bibfield  {title} {\bibinfo {title} {The unpredictability of
  quantum gravity},\ }\href {https://doi.org/10.1007/BF01206031} {\bibfield
  {journal} {\bibinfo  {journal} {Communications in Mathematical Physics}\
  }\textbf {\bibinfo {volume} {87}},\ \bibinfo {pages} {395} (\bibinfo {year}
  {1982})}\BibitemShut {NoStop}%
\bibitem [{\citenamefont {Wheeler}(1955)}]{PhysRev.97.511}%
  \BibitemOpen
  \bibfield  {author} {\bibinfo {author} {\bibfnamefont {J.~A.}\ \bibnamefont
  {Wheeler}},\ }\bibfield  {title} {\bibinfo {title} {Geons},\ }\href
  {https://doi.org/10.1103/PhysRev.97.511} {\bibfield  {journal} {\bibinfo
  {journal} {Phys. Rev.}\ }\textbf {\bibinfo {volume} {97}},\ \bibinfo {pages}
  {511} (\bibinfo {year} {1955})}\BibitemShut {NoStop}%
\bibitem [{\citenamefont {Coelho}\ and\ \citenamefont
  {Mann}(2017)}]{PhysRevD.96.093009}%
  \BibitemOpen
  \bibfield  {author} {\bibinfo {author} {\bibfnamefont {J.~a. A.~B.}\
  \bibnamefont {Coelho}}\ and\ \bibinfo {author} {\bibfnamefont {W.~A.}\
  \bibnamefont {Mann}},\ }\bibfield  {title} {\bibinfo {title} {Decoherence,
  matter effect, and neutrino hierarchy signature in long baseline
  experiments},\ }\href {https://doi.org/10.1103/PhysRevD.96.093009} {\bibfield
   {journal} {\bibinfo  {journal} {Phys. Rev. D}\ }\textbf {\bibinfo {volume}
  {96}},\ \bibinfo {pages} {093009} (\bibinfo {year} {2017})}\BibitemShut
  {NoStop}%
\bibitem [{\citenamefont {Buoninfante}\ \emph {et~al.}(2020)\citenamefont
  {Buoninfante}, \citenamefont {Capolupo}, \citenamefont {Giampaolo},\ and\
  \citenamefont {Lambiase}}]{Buoninfante:2020iyr}%
  \BibitemOpen
  \bibfield  {author} {\bibinfo {author} {\bibfnamefont {L.}~\bibnamefont
  {Buoninfante}}, \bibinfo {author} {\bibfnamefont {A.}~\bibnamefont
  {Capolupo}}, \bibinfo {author} {\bibfnamefont {S.~M.}\ \bibnamefont
  {Giampaolo}},\ and\ \bibinfo {author} {\bibfnamefont {G.}~\bibnamefont
  {Lambiase}},\ }\bibfield  {title} {\bibinfo {title} {{Revealing neutrino
  nature and $CPT$ violation with decoherence effects}},\ }\href@noop {} {\
  (\bibinfo {year} {2020})},\ \Eprint {https://arxiv.org/abs/2001.07580}
  {arXiv:2001.07580 [hep-ph]} \BibitemShut {NoStop}%
\bibitem [{\citenamefont {Carrasco}\ \emph {et~al.}(2019)\citenamefont
  {Carrasco}, \citenamefont {D\'{\i}az},\ and\ \citenamefont
  {Gago}}]{PhysRevD.99.075022}%
  \BibitemOpen
  \bibfield  {author} {\bibinfo {author} {\bibfnamefont {J.~C.}\ \bibnamefont
  {Carrasco}}, \bibinfo {author} {\bibfnamefont {F.~N.}\ \bibnamefont
  {D\'{\i}az}},\ and\ \bibinfo {author} {\bibfnamefont {A.~M.}\ \bibnamefont
  {Gago}},\ }\bibfield  {title} {\bibinfo {title} {{Probing $CPT$ breaking
  induced by quantum decoherence at DUNE}},\ }\href
  {https://doi.org/10.1103/PhysRevD.99.075022} {\bibfield  {journal} {\bibinfo
  {journal} {Phys. Rev. D}\ }\textbf {\bibinfo {volume} {99}},\ \bibinfo
  {pages} {075022} (\bibinfo {year} {2019})}\BibitemShut {NoStop}%
\bibitem [{\citenamefont {Lisi}\ \emph {et~al.}(2000)\citenamefont {Lisi},
  \citenamefont {Marrone},\ and\ \citenamefont
  {Montanino}}]{PhysRevLett.85.1166}%
  \BibitemOpen
  \bibfield  {author} {\bibinfo {author} {\bibfnamefont {E.}~\bibnamefont
  {Lisi}}, \bibinfo {author} {\bibfnamefont {A.}~\bibnamefont {Marrone}},\ and\
  \bibinfo {author} {\bibfnamefont {D.}~\bibnamefont {Montanino}},\ }\bibfield
  {title} {\bibinfo {title} {Probing possible decoherence effects in
  atmospheric neutrino oscillations},\ }\href
  {https://doi.org/10.1103/PhysRevLett.85.1166} {\bibfield  {journal} {\bibinfo
   {journal} {Phys. Rev. Lett.}\ }\textbf {\bibinfo {volume} {85}},\ \bibinfo
  {pages} {1166} (\bibinfo {year} {2000})}\BibitemShut {NoStop}%
\bibitem [{\citenamefont {Ohlsson}(2001)}]{OHLSSON2001159}%
  \BibitemOpen
  \bibfield  {author} {\bibinfo {author} {\bibfnamefont {T.}~\bibnamefont
  {Ohlsson}},\ }\bibfield  {title} {\bibinfo {title} {Equivalence between
  gaussian averaged neutrino oscillations and neutrino decoherence},\ }\href
  {https://doi.org/https://doi.org/10.1016/S0370-2693(01)00178-2} {\bibfield
  {journal} {\bibinfo  {journal} {Physics Letters B}\ }\textbf {\bibinfo
  {volume} {502}},\ \bibinfo {pages} {159 } (\bibinfo {year}
  {2001})}\BibitemShut {NoStop}%
\bibitem [{\citenamefont {Farzan}\ \emph {et~al.}(2008)\citenamefont {Farzan},
  \citenamefont {Schwetz},\ and\ \citenamefont {Smirnov}}]{Farzan:2008zv}%
  \BibitemOpen
  \bibfield  {author} {\bibinfo {author} {\bibfnamefont {Y.}~\bibnamefont
  {Farzan}}, \bibinfo {author} {\bibfnamefont {T.}~\bibnamefont {Schwetz}},\
  and\ \bibinfo {author} {\bibfnamefont {A.~Y.}\ \bibnamefont {Smirnov}},\
  }\bibfield  {title} {\bibinfo {title} {{Reconciling results of LSND,
  MiniBooNE and other experiments with soft decoherence}},\ }\href
  {https://doi.org/10.1088/1126-6708/2008/07/067} {\bibfield  {journal}
  {\bibinfo  {journal} {JHEP}\ }\textbf {\bibinfo {volume} {07}},\ \bibinfo
  {pages} {067}},\ \Eprint {https://arxiv.org/abs/0805.2098} {arXiv:0805.2098
  [hep-ph]} \BibitemShut {NoStop}%
\bibitem [{\citenamefont {Coloma}\ \emph {et~al.}(2018)\citenamefont {Coloma},
  \citenamefont {Lopez-Pavon}, \citenamefont {Martinez-Soler},\ and\
  \citenamefont {Nunokawa}}]{Coloma:2018idr}%
  \BibitemOpen
  \bibfield  {author} {\bibinfo {author} {\bibfnamefont {P.}~\bibnamefont
  {Coloma}}, \bibinfo {author} {\bibfnamefont {J.}~\bibnamefont {Lopez-Pavon}},
  \bibinfo {author} {\bibfnamefont {I.}~\bibnamefont {Martinez-Soler}},\ and\
  \bibinfo {author} {\bibfnamefont {H.}~\bibnamefont {Nunokawa}},\ }\bibfield
  {title} {\bibinfo {title} {{Decoherence in Neutrino Propagation Through
  Matter, and Bounds from IceCube/DeepCore}},\ }\href
  {https://doi.org/10.1140/epjc/s10052-018-6092-6} {\bibfield  {journal}
  {\bibinfo  {journal} {Eur. Phys. J.}\ }\textbf {\bibinfo {volume} {C78}},\
  \bibinfo {pages} {614} (\bibinfo {year} {2018})},\ \Eprint
  {https://arxiv.org/abs/1803.04438} {arXiv:1803.04438 [hep-ph]} \BibitemShut
  {NoStop}%
\bibitem [{\citenamefont {Carpio}\ \emph {et~al.}(2019)\citenamefont {Carpio},
  \citenamefont {Massoni},\ and\ \citenamefont {Gago}}]{Carpio:2018gum}%
  \BibitemOpen
  \bibfield  {author} {\bibinfo {author} {\bibfnamefont {J.~A.}\ \bibnamefont
  {Carpio}}, \bibinfo {author} {\bibfnamefont {E.}~\bibnamefont {Massoni}},\
  and\ \bibinfo {author} {\bibfnamefont {A.~M.}\ \bibnamefont {Gago}},\
  }\bibfield  {title} {\bibinfo {title} {{Testing quantum decoherence at
  DUNE}},\ }\href {https://doi.org/10.1103/PhysRevD.100.015035} {\bibfield
  {journal} {\bibinfo  {journal} {Phys. Rev.}\ }\textbf {\bibinfo {volume}
  {D100}},\ \bibinfo {pages} {015035} (\bibinfo {year} {2019})},\ \Eprint
  {https://arxiv.org/abs/1811.07923} {arXiv:1811.07923 [hep-ph]} \BibitemShut
  {NoStop}%
\bibitem [{\citenamefont {Carpio}\ \emph {et~al.}(2018)\citenamefont {Carpio},
  \citenamefont {Massoni},\ and\ \citenamefont {Gago}}]{Carpio:2017nui}%
  \BibitemOpen
  \bibfield  {author} {\bibinfo {author} {\bibfnamefont {J.~A.}\ \bibnamefont
  {Carpio}}, \bibinfo {author} {\bibfnamefont {E.}~\bibnamefont {Massoni}},\
  and\ \bibinfo {author} {\bibfnamefont {A.~M.}\ \bibnamefont {Gago}},\
  }\bibfield  {title} {\bibinfo {title} {{Revisiting quantum decoherence for
  neutrino oscillations in matter with constant density}},\ }\href
  {https://doi.org/10.1103/PhysRevD.97.115017} {\bibfield  {journal} {\bibinfo
  {journal} {Phys. Rev.}\ }\textbf {\bibinfo {volume} {D97}},\ \bibinfo {pages}
  {115017} (\bibinfo {year} {2018})},\ \Eprint
  {https://arxiv.org/abs/1711.03680} {arXiv:1711.03680 [hep-ph]} \BibitemShut
  {NoStop}%
\bibitem [{\citenamefont {Anchordoqui}\ \emph {et~al.}(2005)\citenamefont
  {Anchordoqui}, \citenamefont {Goldberg}, \citenamefont {Gonzalez-Garcia},
  \citenamefont {Halzen}, \citenamefont {Hooper}, \citenamefont {Sarkar},\ and\
  \citenamefont {Weiler}}]{Anchordoqui:2005gj}%
  \BibitemOpen
  \bibfield  {author} {\bibinfo {author} {\bibfnamefont {L.~A.}\ \bibnamefont
  {Anchordoqui}}, \bibinfo {author} {\bibfnamefont {H.}~\bibnamefont
  {Goldberg}}, \bibinfo {author} {\bibfnamefont {M.~C.}\ \bibnamefont
  {Gonzalez-Garcia}}, \bibinfo {author} {\bibfnamefont {F.}~\bibnamefont
  {Halzen}}, \bibinfo {author} {\bibfnamefont {D.}~\bibnamefont {Hooper}},
  \bibinfo {author} {\bibfnamefont {S.}~\bibnamefont {Sarkar}},\ and\ \bibinfo
  {author} {\bibfnamefont {T.~J.}\ \bibnamefont {Weiler}},\ }\bibfield  {title}
  {\bibinfo {title} {{Probing Planck scale physics with IceCube}},\ }\href
  {https://doi.org/10.1103/PhysRevD.72.065019} {\bibfield  {journal} {\bibinfo
  {journal} {Phys. Rev.}\ }\textbf {\bibinfo {volume} {D72}},\ \bibinfo {pages}
  {065019} (\bibinfo {year} {2005})},\ \Eprint
  {https://arxiv.org/abs/hep-ph/0506168} {arXiv:hep-ph/0506168 [hep-ph]}
  \BibitemShut {NoStop}%
\bibitem [{\citenamefont {Balieiro~Gomes}\ \emph {et~al.}(2017)\citenamefont
  {Balieiro~Gomes}, \citenamefont {Guzzo}, \citenamefont {de~Holanda},\ and\
  \citenamefont {Oliveira}}]{PhysRevD.95.113005}%
  \BibitemOpen
  \bibfield  {author} {\bibinfo {author} {\bibfnamefont {G.}~\bibnamefont
  {Balieiro~Gomes}}, \bibinfo {author} {\bibfnamefont {M.~M.}\ \bibnamefont
  {Guzzo}}, \bibinfo {author} {\bibfnamefont {P.~C.}\ \bibnamefont
  {de~Holanda}},\ and\ \bibinfo {author} {\bibfnamefont {R.~L.~N.}\
  \bibnamefont {Oliveira}},\ }\bibfield  {title} {\bibinfo {title} {Parameter
  limits for neutrino oscillation with decoherence in kamland},\ }\href
  {https://doi.org/10.1103/PhysRevD.95.113005} {\bibfield  {journal} {\bibinfo
  {journal} {Phys. Rev. D}\ }\textbf {\bibinfo {volume} {95}},\ \bibinfo
  {pages} {113005} (\bibinfo {year} {2017})}\BibitemShut {NoStop}%
\bibitem [{\citenamefont {Jones}(2015)}]{PhysRevD.91.053002}%
  \BibitemOpen
  \bibfield  {author} {\bibinfo {author} {\bibfnamefont {B.~J.~P.}\
  \bibnamefont {Jones}},\ }\bibfield  {title} {\bibinfo {title} {Dynamical pion
  collapse and the coherence of conventional neutrino beams},\ }\href
  {https://doi.org/10.1103/PhysRevD.91.053002} {\bibfield  {journal} {\bibinfo
  {journal} {Phys. Rev. D}\ }\textbf {\bibinfo {volume} {91}},\ \bibinfo
  {pages} {053002} (\bibinfo {year} {2015})}\BibitemShut {NoStop}%
\bibitem [{\citenamefont {Fogli}\ \emph {et~al.}(2007)\citenamefont {Fogli},
  \citenamefont {Lisi}, \citenamefont {Marrone}, \citenamefont {Montanino},\
  and\ \citenamefont {Palazzo}}]{PhysRevD.76.033006}%
  \BibitemOpen
  \bibfield  {author} {\bibinfo {author} {\bibfnamefont {G.~L.}\ \bibnamefont
  {Fogli}}, \bibinfo {author} {\bibfnamefont {E.}~\bibnamefont {Lisi}},
  \bibinfo {author} {\bibfnamefont {A.}~\bibnamefont {Marrone}}, \bibinfo
  {author} {\bibfnamefont {D.}~\bibnamefont {Montanino}},\ and\ \bibinfo
  {author} {\bibfnamefont {A.}~\bibnamefont {Palazzo}},\ }\bibfield  {title}
  {\bibinfo {title} {Probing nonstandard decoherence effects with solar and
  kamland neutrinos},\ }\href {https://doi.org/10.1103/PhysRevD.76.033006}
  {\bibfield  {journal} {\bibinfo  {journal} {Phys. Rev. D}\ }\textbf {\bibinfo
  {volume} {76}},\ \bibinfo {pages} {033006} (\bibinfo {year}
  {2007})}\BibitemShut {NoStop}%
\bibitem [{\citenamefont {Coelho}\ \emph {et~al.}(2017)\citenamefont {Coelho},
  \citenamefont {Mann},\ and\ \citenamefont {Bashar}}]{PhysRevLett.118.221801}%
  \BibitemOpen
  \bibfield  {author} {\bibinfo {author} {\bibfnamefont {J.~a. A.~B.}\
  \bibnamefont {Coelho}}, \bibinfo {author} {\bibfnamefont {W.~A.}\
  \bibnamefont {Mann}},\ and\ \bibinfo {author} {\bibfnamefont {S.~S.}\
  \bibnamefont {Bashar}},\ }\bibfield  {title} {\bibinfo {title} {Nonmaximal
  ${\ensuremath{\theta}}_{23}$ mixing at nova from neutrino decoherence},\
  }\href {https://doi.org/10.1103/PhysRevLett.118.221801} {\bibfield  {journal}
  {\bibinfo  {journal} {Phys. Rev. Lett.}\ }\textbf {\bibinfo {volume} {118}},\
  \bibinfo {pages} {221801} (\bibinfo {year} {2017})}\BibitemShut {NoStop}%
\bibitem [{\citenamefont {Guzzo}\ \emph {et~al.}(2016)\citenamefont {Guzzo},
  \citenamefont {de~Holanda},\ and\ \citenamefont {Oliveira}}]{GUZZO2016408}%
  \BibitemOpen
  \bibfield  {author} {\bibinfo {author} {\bibfnamefont {M.~M.}\ \bibnamefont
  {Guzzo}}, \bibinfo {author} {\bibfnamefont {P.~C.}\ \bibnamefont
  {de~Holanda}},\ and\ \bibinfo {author} {\bibfnamefont {R.~L.}\ \bibnamefont
  {Oliveira}},\ }\bibfield  {title} {\bibinfo {title} {Quantum dissipation in a
  neutrino system propagating in vacuum and in matter},\ }\href
  {https://doi.org/https://doi.org/10.1016/j.nuclphysb.2016.04.030} {\bibfield
  {journal} {\bibinfo  {journal} {Nuclear Physics B}\ }\textbf {\bibinfo
  {volume} {908}},\ \bibinfo {pages} {408 } (\bibinfo {year} {2016})},\
  \bibinfo {note} {neutrino Oscillations: Celebrating the Nobel Prize in
  Physics 2015}\BibitemShut {NoStop}%
\bibitem [{\citenamefont {Morgan}\ \emph {et~al.}(2006)\citenamefont {Morgan},
  \citenamefont {Winstanley}, \citenamefont {Brunner},\ and\ \citenamefont
  {Thompson}}]{Morgan:2004vv}%
  \BibitemOpen
  \bibfield  {author} {\bibinfo {author} {\bibfnamefont {D.}~\bibnamefont
  {Morgan}}, \bibinfo {author} {\bibfnamefont {E.}~\bibnamefont {Winstanley}},
  \bibinfo {author} {\bibfnamefont {J.}~\bibnamefont {Brunner}},\ and\ \bibinfo
  {author} {\bibfnamefont {L.~F.}\ \bibnamefont {Thompson}},\ }\bibfield
  {title} {\bibinfo {title} {{Probing quantum decoherence in atmospheric
  neutrino oscillations with a neutrino telescope}},\ }\href
  {https://doi.org/10.1016/j.astropartphys.2006.03.001} {\bibfield  {journal}
  {\bibinfo  {journal} {Astropart. Phys.}\ }\textbf {\bibinfo {volume} {25}},\
  \bibinfo {pages} {311} (\bibinfo {year} {2006})},\ \Eprint
  {https://arxiv.org/abs/astro-ph/0412618} {arXiv:astro-ph/0412618 [astro-ph]}
  \BibitemShut {NoStop}%
\bibitem [{\citenamefont {Abbasi}\ \emph {et~al.}(2009)\citenamefont {Abbasi}
  \emph {et~al.}}]{Abbasi:2009nfa}%
  \BibitemOpen
  \bibfield  {author} {\bibinfo {author} {\bibfnamefont {R.}~\bibnamefont
  {Abbasi}} \emph {et~al.} (\bibinfo {collaboration} {IceCube}),\ }\bibfield
  {title} {\bibinfo {title} {{Determination of the Atmospheric Neutrino Flux
  and Searches for New Physics with AMANDA-II}},\ }\href
  {https://doi.org/10.1103/PhysRevD.79.102005} {\bibfield  {journal} {\bibinfo
  {journal} {Phys. Rev.}\ }\textbf {\bibinfo {volume} {D79}},\ \bibinfo {pages}
  {102005} (\bibinfo {year} {2009})},\ \Eprint
  {https://arxiv.org/abs/0902.0675} {arXiv:0902.0675 [astro-ph.HE]}
  \BibitemShut {NoStop}%
\bibitem [{\citenamefont {Gomes}\ \emph {et~al.}(2020)\citenamefont {Gomes},
  \citenamefont {Gomes},\ and\ \citenamefont {Peres}}]{Gomes:2020muc}%
  \BibitemOpen
  \bibfield  {author} {\bibinfo {author} {\bibfnamefont {A.}~\bibnamefont
  {Gomes}}, \bibinfo {author} {\bibfnamefont {R.}~\bibnamefont {Gomes}},\ and\
  \bibinfo {author} {\bibfnamefont {O.}~\bibnamefont {Peres}},\ }\bibfield
  {title} {\bibinfo {title} {{Quantum decoherence and relaxation in neutrinos
  using long-baseline data}},\ }\href@noop {} {\  (\bibinfo {year} {2020})},\
  \Eprint {https://arxiv.org/abs/2001.09250} {arXiv:2001.09250 [hep-ph]}
  \BibitemShut {NoStop}%
\bibitem [{\citenamefont {Ohlsson}\ and\ \citenamefont
  {Zhou}(2020)}]{Ohlsson:2020gxx}%
  \BibitemOpen
  \bibfield  {author} {\bibinfo {author} {\bibfnamefont {T.}~\bibnamefont
  {Ohlsson}}\ and\ \bibinfo {author} {\bibfnamefont {S.}~\bibnamefont {Zhou}},\
  }\bibfield  {title} {\bibinfo {title} {{Density Matrix Formalism for
  PT-Symmetric Non-Hermitian Hamiltonians with the Lindblad Equation}},\
  }\href@noop {} {\  (\bibinfo {year} {2020})},\ \Eprint
  {https://arxiv.org/abs/2006.02445} {arXiv:2006.02445 [quant-ph]} \BibitemShut
  {NoStop}%
\bibitem [{\citenamefont {Pontecorvo}(1958)}]{Pontecorvo:1957qd}%
  \BibitemOpen
  \bibfield  {author} {\bibinfo {author} {\bibfnamefont {B.}~\bibnamefont
  {Pontecorvo}},\ }\bibfield  {title} {\bibinfo {title} {Inverse beta processes
  and nonconservation of lepton charge},\ }\href@noop {} {\bibfield  {journal}
  {\bibinfo  {journal} {Sov. Phys. JETP}\ }\textbf {\bibinfo {volume} {7}},\
  \bibinfo {pages} {172} (\bibinfo {year} {1958})}\BibitemShut {NoStop}%
\bibitem [{\citenamefont {Maki}\ \emph {et~al.}(1962)\citenamefont {Maki},
  \citenamefont {Nakagawa},\ and\ \citenamefont {Sakata}}]{Maki:1962mu}%
  \BibitemOpen
  \bibfield  {author} {\bibinfo {author} {\bibfnamefont {Z.}~\bibnamefont
  {Maki}}, \bibinfo {author} {\bibfnamefont {M.}~\bibnamefont {Nakagawa}},\
  and\ \bibinfo {author} {\bibfnamefont {S.}~\bibnamefont {Sakata}},\
  }\bibfield  {title} {\bibinfo {title} {{Remarks on the unified model of
  elementary particles}},\ }\href {https://doi.org/10.1143/PTP.28.870}
  {\bibfield  {journal} {\bibinfo  {journal} {Prog. Theor. Phys.}\ }\textbf
  {\bibinfo {volume} {28}},\ \bibinfo {pages} {870} (\bibinfo {year}
  {1962})}\BibitemShut {NoStop}%
\bibitem [{\citenamefont {Ellis}\ \emph {et~al.}(1992)\citenamefont {Ellis},
  \citenamefont {Mavromatos},\ and\ \citenamefont {Nanopoulos}}]{ELLIS199237}%
  \BibitemOpen
  \bibfield  {author} {\bibinfo {author} {\bibfnamefont {J.}~\bibnamefont
  {Ellis}}, \bibinfo {author} {\bibfnamefont {N.}~\bibnamefont {Mavromatos}},\
  and\ \bibinfo {author} {\bibfnamefont {D.}~\bibnamefont {Nanopoulos}},\
  }\bibfield  {title} {\bibinfo {title} {String theory modifies quantum
  mechanics},\ }\href
  {https://doi.org/https://doi.org/10.1016/0370-2693(92)91478-R} {\bibfield
  {journal} {\bibinfo  {journal} {Physics Letters B}\ }\textbf {\bibinfo
  {volume} {293}},\ \bibinfo {pages} {37 } (\bibinfo {year}
  {1992})}\BibitemShut {NoStop}%
\bibitem [{\citenamefont {Misner}\ \emph {et~al.}(1973)\citenamefont {Misner},
  \citenamefont {Thorne},\ and\ \citenamefont
  {Wheeler}}]{misner1973gravitation}%
  \BibitemOpen
  \bibfield  {author} {\bibinfo {author} {\bibfnamefont {C.}~\bibnamefont
  {Misner}}, \bibinfo {author} {\bibfnamefont {K.}~\bibnamefont {Thorne}},\
  and\ \bibinfo {author} {\bibfnamefont {J.}~\bibnamefont {Wheeler}},\
  }\href@noop {} {\emph {\bibinfo {title} {Gravitation}}}\ (\bibinfo
  {publisher} {W. H. Freeman},\ \bibinfo {year} {1973})\BibitemShut {NoStop}%
\bibitem [{\citenamefont {Ng}(2010)}]{ng2010holographic}%
  \BibitemOpen
  \bibfield  {author} {\bibinfo {author} {\bibfnamefont {Y.~J.}\ \bibnamefont
  {Ng}},\ }\href@noop {} {\bibinfo {title} {Holographic quantum foam}}
  (\bibinfo {year} {2010}),\ \Eprint {https://arxiv.org/abs/1001.0411}
  {arXiv:1001.0411 [gr-qc]} \BibitemShut {NoStop}%
\bibitem [{\citenamefont {Pauli}(1956)}]{PauliLightcone}%
  \BibitemOpen
  \bibfield  {author} {\bibinfo {author} {\bibfnamefont {W.}~\bibnamefont
  {Pauli}},\ }\href@noop {} {\bibfield  {journal} {\bibinfo  {journal} {Helv.
  Phys. Acta. Suppl. 4, No. 69.}\ } (\bibinfo {year} {1956})}\BibitemShut
  {NoStop}%
\bibitem [{\citenamefont {Ford}(1995)}]{Ford_1995}%
  \BibitemOpen
  \bibfield  {author} {\bibinfo {author} {\bibfnamefont {L.~H.}\ \bibnamefont
  {Ford}},\ }\bibfield  {title} {\bibinfo {title} {Gravitons and light cone
  fluctuations},\ }\href {https://doi.org/10.1103/physrevd.51.1692} {\bibfield
  {journal} {\bibinfo  {journal} {Physical Review D}\ }\textbf {\bibinfo
  {volume} {51}},\ \bibinfo {pages} {1692–1700} (\bibinfo {year}
  {1995})}\BibitemShut {NoStop}%
\bibitem [{\citenamefont {Yu}\ \emph {et~al.}(2009)\citenamefont {Yu},
  \citenamefont {Svaiter},\ and\ \citenamefont {Ford}}]{Yu_2009}%
  \BibitemOpen
  \bibfield  {author} {\bibinfo {author} {\bibfnamefont {H.}~\bibnamefont
  {Yu}}, \bibinfo {author} {\bibfnamefont {N.~F.}\ \bibnamefont {Svaiter}},\
  and\ \bibinfo {author} {\bibfnamefont {L.~H.}\ \bibnamefont {Ford}},\
  }\bibfield  {title} {\bibinfo {title} {Quantum light-cone fluctuations in
  compactified spacetimes},\ }\bibfield  {journal} {\bibinfo  {journal}
  {Physical Review D}\ }\textbf {\bibinfo {volume} {80}},\ \href
  {https://doi.org/10.1103/physrevd.80.124019} {10.1103/physrevd.80.124019}
  (\bibinfo {year} {2009})\BibitemShut {NoStop}%
\bibitem [{\citenamefont {Barrow}(1987)}]{BARROW198712}%
  \BibitemOpen
  \bibfield  {author} {\bibinfo {author} {\bibfnamefont {J.~D.}\ \bibnamefont
  {Barrow}},\ }\bibfield  {title} {\bibinfo {title} {Cosmic no-hair theorems
  and inflation},\ }\href
  {https://doi.org/https://doi.org/10.1016/0370-2693(87)90063-3} {\bibfield
  {journal} {\bibinfo  {journal} {Physics Letters B}\ }\textbf {\bibinfo
  {volume} {187}},\ \bibinfo {pages} {12 } (\bibinfo {year}
  {1987})}\BibitemShut {NoStop}%
\bibitem [{\citenamefont {Adams}\ \emph {et~al.}(2001)\citenamefont {Adams},
  \citenamefont {Kane}, \citenamefont {Mbonye},\ and\ \citenamefont
  {Perry}}]{ProtonDecay}%
  \BibitemOpen
  \bibfield  {author} {\bibinfo {author} {\bibfnamefont {F.~C.}\ \bibnamefont
  {Adams}}, \bibinfo {author} {\bibfnamefont {G.~L.}\ \bibnamefont {Kane}},
  \bibinfo {author} {\bibfnamefont {M.}~\bibnamefont {Mbonye}},\ and\ \bibinfo
  {author} {\bibfnamefont {M.~J.}\ \bibnamefont {Perry}},\ }\bibfield  {title}
  {\bibinfo {title} {Proton decay, black holes, and large extra dimensions},\
  }\href {https://doi.org/10.1142/S0217751X0100369X} {\bibfield  {journal}
  {\bibinfo  {journal} {International Journal of Modern Physics A}\ }\textbf
  {\bibinfo {volume} {16}},\ \bibinfo {pages} {2399} (\bibinfo {year}
  {2001})}\BibitemShut {NoStop}%
\bibitem [{\citenamefont {Alsaleh}\ \emph {et~al.}(2017)\citenamefont
  {Alsaleh}, \citenamefont {Al-Modlej},\ and\ \citenamefont
  {Ali}}]{Alsaleh_2017}%
  \BibitemOpen
  \bibfield  {author} {\bibinfo {author} {\bibfnamefont {S.}~\bibnamefont
  {Alsaleh}}, \bibinfo {author} {\bibfnamefont {A.}~\bibnamefont {Al-Modlej}},\
  and\ \bibinfo {author} {\bibfnamefont {A.~F.}\ \bibnamefont {Ali}},\
  }\bibfield  {title} {\bibinfo {title} {Virtual black holes from the
  generalized uncertainty principle and proton decay},\ }\href
  {https://doi.org/10.1209/0295-5075/118/50008} {\bibfield  {journal} {\bibinfo
   {journal} {{EPL} (Europhysics Letters)}\ }\textbf {\bibinfo {volume}
  {118}},\ \bibinfo {pages} {50008} (\bibinfo {year} {2017})}\BibitemShut
  {NoStop}%
\bibitem [{\citenamefont {Perlman}\ \emph {et~al.}(2015)\citenamefont
  {Perlman}, \citenamefont {Rappaport}, \citenamefont {Christiansen},
  \citenamefont {Ng}, \citenamefont {DeVore},\ and\ \citenamefont
  {Pooley}}]{Perlman_2015}%
  \BibitemOpen
  \bibfield  {author} {\bibinfo {author} {\bibfnamefont {E.~S.}\ \bibnamefont
  {Perlman}}, \bibinfo {author} {\bibfnamefont {S.~A.}\ \bibnamefont
  {Rappaport}}, \bibinfo {author} {\bibfnamefont {W.~A.}\ \bibnamefont
  {Christiansen}}, \bibinfo {author} {\bibfnamefont {Y.~J.}\ \bibnamefont
  {Ng}}, \bibinfo {author} {\bibfnamefont {J.}~\bibnamefont {DeVore}},\ and\
  \bibinfo {author} {\bibfnamefont {D.}~\bibnamefont {Pooley}},\ }\bibfield
  {title} {\bibinfo {title} {New constraints on quantum gravity from x-ray and
  gamma-ray observations},\ }\href {https://doi.org/10.1088/0004-637x/805/1/10}
  {\bibfield  {journal} {\bibinfo  {journal} {The Astrophysical Journal}\
  }\textbf {\bibinfo {volume} {805}},\ \bibinfo {pages} {10} (\bibinfo {year}
  {2015})}\BibitemShut {NoStop}%
\bibitem [{\citenamefont {Chatelain}\ and\ \citenamefont
  {Volpe}(2020)}]{CHATELAIN2020135150}%
  \BibitemOpen
  \bibfield  {author} {\bibinfo {author} {\bibfnamefont {A.}~\bibnamefont
  {Chatelain}}\ and\ \bibinfo {author} {\bibfnamefont {M.~C.}\ \bibnamefont
  {Volpe}},\ }\bibfield  {title} {\bibinfo {title} {Neutrino decoherence in
  presence of strong gravitational fields},\ }\href
  {https://doi.org/https://doi.org/10.1016/j.physletb.2019.135150} {\bibfield
  {journal} {\bibinfo  {journal} {Physics Letters B}\ }\textbf {\bibinfo
  {volume} {801}},\ \bibinfo {pages} {135150} (\bibinfo {year}
  {2020})}\BibitemShut {NoStop}%
\bibitem [{\citenamefont {Dvornikov}(2019)}]{PhysRevD.100.096014}%
  \BibitemOpen
  \bibfield  {author} {\bibinfo {author} {\bibfnamefont {M.}~\bibnamefont
  {Dvornikov}},\ }\bibfield  {title} {\bibinfo {title} {Neutrino flavor
  oscillations in stochastic gravitational waves},\ }\href
  {https://doi.org/10.1103/PhysRevD.100.096014} {\bibfield  {journal} {\bibinfo
   {journal} {Phys. Rev. D}\ }\textbf {\bibinfo {volume} {100}},\ \bibinfo
  {pages} {096014} (\bibinfo {year} {2019})}\BibitemShut {NoStop}%
\bibitem [{\citenamefont {Jacobsson}\ \emph {et~al.}(2002)\citenamefont
  {Jacobsson}, \citenamefont {Ohlsson}, \citenamefont {Snellman},\ and\
  \citenamefont {Winter}}]{MatterDensityFluctuation}%
  \BibitemOpen
  \bibfield  {author} {\bibinfo {author} {\bibfnamefont {B.}~\bibnamefont
  {Jacobsson}}, \bibinfo {author} {\bibfnamefont {T.}~\bibnamefont {Ohlsson}},
  \bibinfo {author} {\bibfnamefont {H.}~\bibnamefont {Snellman}},\ and\
  \bibinfo {author} {\bibfnamefont {W.}~\bibnamefont {Winter}},\ }\bibfield
  {title} {\bibinfo {title} {Effects of random matter density fluctuations on
  the neutrino oscillation transition probabilities in the earth},\ }\href
  {https://doi.org/10.1016/S0370-2693(02)01580-0} {\bibfield  {journal}
  {\bibinfo  {journal} {Physics Letters B}\ }\textbf {\bibinfo {volume}
  {532}},\ \bibinfo {pages} {259} (\bibinfo {year} {2002})}\BibitemShut
  {NoStop}%
\bibitem [{\citenamefont {Nieves}\ and\ \citenamefont
  {Sahu}(2020)}]{2002.08315}%
  \BibitemOpen
  \bibfield  {author} {\bibinfo {author} {\bibfnamefont {J.~F.}\ \bibnamefont
  {Nieves}}\ and\ \bibinfo {author} {\bibfnamefont {S.}~\bibnamefont {Sahu}},\
  }\href@noop {} {\bibinfo {title} {Neutrino decoherence in an electron and
  nucleon background}} (\bibinfo {year} {2020}),\ \Eprint
  {https://arxiv.org/abs/arXiv:2002.08315} {arXiv:2002.08315} \BibitemShut
  {NoStop}%
\bibitem [{\citenamefont {Lindner}\ \emph {et~al.}(2001)\citenamefont
  {Lindner}, \citenamefont {Ohlsson},\ and\ \citenamefont
  {Winter}}]{LINDNER2001326}%
  \BibitemOpen
  \bibfield  {author} {\bibinfo {author} {\bibfnamefont {M.}~\bibnamefont
  {Lindner}}, \bibinfo {author} {\bibfnamefont {T.}~\bibnamefont {Ohlsson}},\
  and\ \bibinfo {author} {\bibfnamefont {W.}~\bibnamefont {Winter}},\
  }\bibfield  {title} {\bibinfo {title} {A combined treatment of neutrino decay
  and neutrino oscillations},\ }\href
  {https://doi.org/https://doi.org/10.1016/S0550-3213(01)00237-1} {\bibfield
  {journal} {\bibinfo  {journal} {Nuclear Physics B}\ }\textbf {\bibinfo
  {volume} {607}},\ \bibinfo {pages} {326 } (\bibinfo {year}
  {2001})}\BibitemShut {NoStop}%
\bibitem [{\citenamefont {Berryman}\ \emph {et~al.}(2015)\citenamefont
  {Berryman}, \citenamefont {de~Gouv\^{e}a}, \citenamefont {Hern\`{a}ndez},\
  and\ \citenamefont {Oliveira}}]{BERRYMAN201574}%
  \BibitemOpen
  \bibfield  {author} {\bibinfo {author} {\bibfnamefont {J.~M.}\ \bibnamefont
  {Berryman}}, \bibinfo {author} {\bibfnamefont {A.}~\bibnamefont
  {de~Gouv\^{e}a}}, \bibinfo {author} {\bibfnamefont {D.}~\bibnamefont
  {Hern\`{a}ndez}},\ and\ \bibinfo {author} {\bibfnamefont {R.~L.}\
  \bibnamefont {Oliveira}},\ }\bibfield  {title} {\bibinfo {title} {Non-unitary
  neutrino propagation from neutrino decay},\ }\href
  {https://doi.org/https://doi.org/10.1016/j.physletb.2015.01.002} {\bibfield
  {journal} {\bibinfo  {journal} {Physics Letters B}\ }\textbf {\bibinfo
  {volume} {742}},\ \bibinfo {pages} {74 } (\bibinfo {year}
  {2015})}\BibitemShut {NoStop}%
\bibitem [{\citenamefont {Lindblad}(1976)}]{lindblad1976}%
  \BibitemOpen
  \bibfield  {author} {\bibinfo {author} {\bibfnamefont {G.}~\bibnamefont
  {Lindblad}},\ }\bibfield  {title} {\bibinfo {title} {On the generators of
  quantum dynamical semigroups},\ }\href
  {https://projecteuclid.org:443/euclid.cmp/1103899849} {\bibfield  {journal}
  {\bibinfo  {journal} {Comm. Math. Phys.}\ }\textbf {\bibinfo {volume} {48}},\
  \bibinfo {pages} {119} (\bibinfo {year} {1976})}\BibitemShut {NoStop}%
\bibitem [{\citenamefont {Benatti}\ and\ \citenamefont
  {Floreanini}(2000)}]{Benatti_2000}%
  \BibitemOpen
  \bibfield  {author} {\bibinfo {author} {\bibfnamefont {F.}~\bibnamefont
  {Benatti}}\ and\ \bibinfo {author} {\bibfnamefont {R.}~\bibnamefont
  {Floreanini}},\ }\bibfield  {title} {\bibinfo {title} {Open system approach
  to neutrino oscillations},\ }\href
  {https://doi.org/10.1088/1126-6708/2000/02/032} {\bibfield  {journal}
  {\bibinfo  {journal} {Journal of High Energy Physics}\ }\textbf {\bibinfo
  {volume} {2000}},\ \bibinfo {pages} {032} (\bibinfo {year}
  {2000})}\BibitemShut {NoStop}%
\bibitem [{\citenamefont {Gago}\ \emph {et~al.}(2002)\citenamefont {Gago},
  \citenamefont {Santos}, \citenamefont {Teves},\ and\ \citenamefont
  {Funchal}}]{gago2002study}%
  \BibitemOpen
  \bibfield  {author} {\bibinfo {author} {\bibfnamefont {A.~M.}\ \bibnamefont
  {Gago}}, \bibinfo {author} {\bibfnamefont {E.~M.}\ \bibnamefont {Santos}},
  \bibinfo {author} {\bibfnamefont {W.~J.~C.}\ \bibnamefont {Teves}},\ and\
  \bibinfo {author} {\bibfnamefont {R.~Z.}\ \bibnamefont {Funchal}},\
  }\href@noop {} {\bibinfo {title} {A study on quantum decoherence phenomena
  with three generations of neutrinos}} (\bibinfo {year} {2002}),\ \Eprint
  {https://arxiv.org/abs/hep-ph/0208166} {arXiv:hep-ph/0208166 [hep-ph]}
  \BibitemShut {NoStop}%
\bibitem [{\citenamefont {Arg{\"{u}}elles~Delgado}\ \emph
  {et~al.}(2015)\citenamefont {Arg{\"{u}}elles~Delgado}, \citenamefont
  {Salvad{\'{o}}},\ and\ \citenamefont {Weaver}}]{Delgado:2014kpa}%
  \BibitemOpen
  \bibfield  {author} {\bibinfo {author} {\bibfnamefont {C.~A.}\ \bibnamefont
  {Arg{\"{u}}elles~Delgado}}, \bibinfo {author} {\bibfnamefont
  {J.}~\bibnamefont {Salvad{\'{o}}}},\ and\ \bibinfo {author} {\bibfnamefont
  {C.~N.}\ \bibnamefont {Weaver}},\ }\bibfield  {title} {\bibinfo {title} {{A
  Simple Quantum Integro-Differential Solver (SQuIDS)}},\ }\href
  {https://doi.org/10.1016/j.cpc.2015.06.022} {\bibfield  {journal} {\bibinfo
  {journal} {Comput. Phys. Commun.}\ }\textbf {\bibinfo {volume} {196}},\
  \bibinfo {pages} {569} (\bibinfo {year} {2015})},\ \Eprint
  {https://arxiv.org/abs/1412.3832} {arXiv:1412.3832 [hep-ph]} \BibitemShut
  {NoStop}%
\bibitem [{\citenamefont {Arg{\"{u}}elles~Delgado}\ \emph
  {et~al.}()\citenamefont {Arg{\"{u}}elles~Delgado}, \citenamefont
  {Salvad{\'{o}}},\ and\ \citenamefont {Weaver}}]{nusquidsGIT}%
  \BibitemOpen
  \bibfield  {author} {\bibinfo {author} {\bibfnamefont {C.~A.}\ \bibnamefont
  {Arg{\"{u}}elles~Delgado}}, \bibinfo {author} {\bibfnamefont
  {J.}~\bibnamefont {Salvad{\'{o}}}},\ and\ \bibinfo {author} {\bibfnamefont
  {C.~N.}\ \bibnamefont {Weaver}},\ }\href
  {{https://github.com/arguelles/nuSQuIDS}} {\bibinfo {title} {{nuSQuIDS}}},\
  \bibinfo {howpublished}
  {\url{https://github.com/arguelles/nuSQuIDS}}\BibitemShut {NoStop}%
\bibitem [{\citenamefont {Aartsen}\ \emph {et~al.}(2013)\citenamefont {Aartsen}
  \emph {et~al.}}]{Aartsen:2013jdh}%
  \BibitemOpen
  \bibfield  {author} {\bibinfo {author} {\bibfnamefont {M.~G.}\ \bibnamefont
  {Aartsen}} \emph {et~al.} (\bibinfo {collaboration} {IceCube}),\ }\bibfield
  {title} {\bibinfo {title} {{Evidence for High-Energy Extraterrestrial
  Neutrinos at the IceCube Detector}},\ }\href
  {https://doi.org/10.1126/science.1242856} {\bibfield  {journal} {\bibinfo
  {journal} {Science}\ }\textbf {\bibinfo {volume} {342}},\ \bibinfo {pages}
  {1242856} (\bibinfo {year} {2013})},\ \Eprint
  {https://arxiv.org/abs/1311.5238} {arXiv:1311.5238 [astro-ph.HE]}
  \BibitemShut {NoStop}%
\bibitem [{\citenamefont {Ellis}\ \emph
  {et~al.}(1997{\natexlab{a}})\citenamefont {Ellis}, \citenamefont
  {Mavromatos}, \citenamefont {Nanopoulos},\ and\ \citenamefont
  {Winstanley}}]{Ellis:1996bz}%
  \BibitemOpen
  \bibfield  {author} {\bibinfo {author} {\bibfnamefont {J.~R.}\ \bibnamefont
  {Ellis}}, \bibinfo {author} {\bibfnamefont {N.~E.}\ \bibnamefont
  {Mavromatos}}, \bibinfo {author} {\bibfnamefont {D.~V.}\ \bibnamefont
  {Nanopoulos}},\ and\ \bibinfo {author} {\bibfnamefont {E.}~\bibnamefont
  {Winstanley}},\ }\bibfield  {title} {\bibinfo {title} {{Quantum decoherence
  in a four-dimensional black hole background}},\ }\href
  {https://doi.org/10.1142/S0217732397000248} {\bibfield  {journal} {\bibinfo
  {journal} {Mod. Phys. Lett.}\ }\textbf {\bibinfo {volume} {A12}},\ \bibinfo
  {pages} {243} (\bibinfo {year} {1997}{\natexlab{a}})},\ \Eprint
  {https://arxiv.org/abs/gr-qc/9602011} {arXiv:gr-qc/9602011 [gr-qc]}
  \BibitemShut {NoStop}%
\bibitem [{\citenamefont {Ellis}\ \emph
  {et~al.}(1997{\natexlab{b}})\citenamefont {Ellis}, \citenamefont
  {Mavromatos},\ and\ \citenamefont {Nanopoulos}}]{Ellis:1997jw}%
  \BibitemOpen
  \bibfield  {author} {\bibinfo {author} {\bibfnamefont {J.~R.}\ \bibnamefont
  {Ellis}}, \bibinfo {author} {\bibfnamefont {N.~E.}\ \bibnamefont
  {Mavromatos}},\ and\ \bibinfo {author} {\bibfnamefont {D.~V.}\ \bibnamefont
  {Nanopoulos}},\ }\bibfield  {title} {\bibinfo {title} {{Quantum decoherence
  in a D foam background}},\ }\href {https://doi.org/10.1142/S0217732397001795}
  {\bibfield  {journal} {\bibinfo  {journal} {Mod. Phys. Lett.}\ }\textbf
  {\bibinfo {volume} {A12}},\ \bibinfo {pages} {1759} (\bibinfo {year}
  {1997}{\natexlab{b}})},\ \Eprint {https://arxiv.org/abs/hep-th/9704169}
  {arXiv:hep-th/9704169 [hep-th]} \BibitemShut {NoStop}%
\bibitem [{\citenamefont {Benatti}\ and\ \citenamefont
  {Floreanini}(1999)}]{BENATTI199958}%
  \BibitemOpen
  \bibfield  {author} {\bibinfo {author} {\bibfnamefont {F.}~\bibnamefont
  {Benatti}}\ and\ \bibinfo {author} {\bibfnamefont {R.}~\bibnamefont
  {Floreanini}},\ }\bibfield  {title} {\bibinfo {title} {Non-standard neutral
  kaon dynamics from infinite statistics},\ }\href
  {https://doi.org/https://doi.org/10.1006/aphy.1998.5896} {\bibfield
  {journal} {\bibinfo  {journal} {Annals of Physics}\ }\textbf {\bibinfo
  {volume} {273}},\ \bibinfo {pages} {58 } (\bibinfo {year}
  {1999})}\BibitemShut {NoStop}%
\bibitem [{\citenamefont {Aartsen}\ \emph
  {et~al.}(2018{\natexlab{a}})\citenamefont {Aartsen} \emph
  {et~al.}}]{IceCube:2018cha}%
  \BibitemOpen
  \bibfield  {author} {\bibinfo {author} {\bibfnamefont {M.~G.}\ \bibnamefont
  {Aartsen}} \emph {et~al.} (\bibinfo {collaboration} {IceCube}),\ }\bibfield
  {title} {\bibinfo {title} {{Neutrino emission from the direction of the
  blazar TXS 0506+056 prior to the IceCube-170922A alert}},\ }\href
  {https://doi.org/10.1126/science.aat2890} {\bibfield  {journal} {\bibinfo
  {journal} {Science}\ }\textbf {\bibinfo {volume} {361}},\ \bibinfo {pages}
  {147} (\bibinfo {year} {2018}{\natexlab{a}})},\ \Eprint
  {https://arxiv.org/abs/1807.08794} {arXiv:1807.08794 [astro-ph.HE]}
  \BibitemShut {NoStop}%
\bibitem [{\citenamefont {Aartsen}\ \emph
  {et~al.}(2018{\natexlab{b}})\citenamefont {Aartsen} \emph
  {et~al.}}]{IceCube:2018dnn}%
  \BibitemOpen
  \bibfield  {author} {\bibinfo {author} {\bibfnamefont {M.~G.}\ \bibnamefont
  {Aartsen}} \emph {et~al.} (\bibinfo {collaboration} {Liverpool Telescope,
  MAGIC, H.E.S.S., AGILE, Kiso, VLA/17B-403, INTEGRAL, Kapteyn, Subaru, HAWC,
  Fermi-LAT, ASAS-SN, VERITAS, Kanata, IceCube, Swift NuSTAR}),\ }\bibfield
  {title} {\bibinfo {title} {{Multimessenger observations of a flaring blazar
  coincident with high-energy neutrino IceCube-170922A}},\ }\href
  {https://doi.org/10.1126/science.aat1378} {\bibfield  {journal} {\bibinfo
  {journal} {Science}\ }\textbf {\bibinfo {volume} {361}},\ \bibinfo {pages}
  {eaat1378} (\bibinfo {year} {2018}{\natexlab{b}})},\ \Eprint
  {https://arxiv.org/abs/1807.08816} {arXiv:1807.08816 [astro-ph.HE]}
  \BibitemShut {NoStop}%
\bibitem [{\citenamefont {Ahlers}\ \emph {et~al.}(2018)\citenamefont {Ahlers},
  \citenamefont {Bustamante},\ and\ \citenamefont {Mu}}]{PhysRevD.98.123023}%
  \BibitemOpen
  \bibfield  {author} {\bibinfo {author} {\bibfnamefont {M.}~\bibnamefont
  {Ahlers}}, \bibinfo {author} {\bibfnamefont {M.}~\bibnamefont {Bustamante}},\
  and\ \bibinfo {author} {\bibfnamefont {S.}~\bibnamefont {Mu}},\ }\bibfield
  {title} {\bibinfo {title} {Unitarity bounds of astrophysical neutrinos},\
  }\href {https://doi.org/10.1103/PhysRevD.98.123023} {\bibfield  {journal}
  {\bibinfo  {journal} {Phys. Rev. D}\ }\textbf {\bibinfo {volume} {98}},\
  \bibinfo {pages} {123023} (\bibinfo {year} {2018})}\BibitemShut {NoStop}%
\bibitem [{\citenamefont {Kelly}\ and\ \citenamefont
  {Machado}(2018)}]{Kelly_2018}%
  \BibitemOpen
  \bibfield  {author} {\bibinfo {author} {\bibfnamefont {K.~J.}\ \bibnamefont
  {Kelly}}\ and\ \bibinfo {author} {\bibfnamefont {P.~A.}\ \bibnamefont
  {Machado}},\ }\bibfield  {title} {\bibinfo {title} {Multimessenger astronomy
  and new neutrino physics},\ }\href
  {https://doi.org/10.1088/1475-7516/2018/10/048} {\bibfield  {journal}
  {\bibinfo  {journal} {Journal of Cosmology and Astroparticle Physics}\
  }\textbf {\bibinfo {volume} {2018}}\bibinfo  {number} { (10)},\ \bibinfo
  {pages} {048}}\BibitemShut {NoStop}%
\bibitem [{\citenamefont {Seckel}\ \emph {et~al.}(1990)\citenamefont {Seckel},
  \citenamefont {Stanev},\ and\ \citenamefont {Gaisser}}]{Seckel:1990pc}%
  \BibitemOpen
\bibfield  {number} {  }\bibfield  {author} {\bibinfo {author} {\bibfnamefont
  {D.}~\bibnamefont {Seckel}}, \bibinfo {author} {\bibfnamefont
  {T.}~\bibnamefont {Stanev}},\ and\ \bibinfo {author} {\bibfnamefont {T.~K.}\
  \bibnamefont {Gaisser}},\ }\bibfield  {title} {\bibinfo {title} {{Signatures
  of cosmic ray interactions on the solar surface}},\ }in\ \href@noop {} {\emph
  {\bibinfo {booktitle} {{Contributions to the 21st international cosmic ray
  conference, Adelaide, Australia, jan 6-19, 1990}}}}\ (\bibinfo {year}
  {1990})\ pp.\ \bibinfo {pages} {463--466}\BibitemShut {NoStop}%
\bibitem [{\citenamefont {Arg\"{u}elles}\ \emph {et~al.}(2017)\citenamefont
  {Arg\"{u}elles}, \citenamefont {de~Wasseige}, \citenamefont {Fedynitch},\
  and\ \citenamefont {Jones}}]{Arg_elles_2017}%
  \BibitemOpen
  \bibfield  {author} {\bibinfo {author} {\bibfnamefont {C.}~\bibnamefont
  {Arg\"{u}elles}}, \bibinfo {author} {\bibfnamefont {G.}~\bibnamefont
  {de~Wasseige}}, \bibinfo {author} {\bibfnamefont {A.}~\bibnamefont
  {Fedynitch}},\ and\ \bibinfo {author} {\bibfnamefont {B.}~\bibnamefont
  {Jones}},\ }\bibfield  {title} {\bibinfo {title} {Solar atmospheric neutrinos
  and the sensitivity floor for solar dark matter annihilation searches},\
  }\href {https://doi.org/10.1088/1475-7516/2017/07/024} {\bibfield  {journal}
  {\bibinfo  {journal} {Journal of Cosmology and Astroparticle Physics}\
  }\textbf {\bibinfo {volume} {2017}}\bibinfo  {number} { (07)},\ \bibinfo
  {pages} {024}}\BibitemShut {NoStop}%
\bibitem [{\citenamefont {Ng}\ \emph {et~al.}(2017)\citenamefont {Ng},
  \citenamefont {Beacom}, \citenamefont {Peter},\ and\ \citenamefont
  {Rott}}]{PhysRevD.96.103006}%
  \BibitemOpen
\bibfield  {number} {  }\bibfield  {author} {\bibinfo {author} {\bibfnamefont
  {K.~C.~Y.}\ \bibnamefont {Ng}}, \bibinfo {author} {\bibfnamefont {J.~F.}\
  \bibnamefont {Beacom}}, \bibinfo {author} {\bibfnamefont {A.~H.~G.}\
  \bibnamefont {Peter}},\ and\ \bibinfo {author} {\bibfnamefont
  {C.}~\bibnamefont {Rott}},\ }\bibfield  {title} {\bibinfo {title} {Solar
  atmospheric neutrinos: A new neutrino floor for dark matter searches},\
  }\href {https://doi.org/10.1103/PhysRevD.96.103006} {\bibfield  {journal}
  {\bibinfo  {journal} {Phys. Rev. D}\ }\textbf {\bibinfo {volume} {96}},\
  \bibinfo {pages} {103006} (\bibinfo {year} {2017})}\BibitemShut {NoStop}%
\bibitem [{\citenamefont {Edsjo}\ \emph {et~al.}(2017)\citenamefont {Edsjo},
  \citenamefont {Elevant}, \citenamefont {Enberg},\ and\ \citenamefont
  {Niblaeus}}]{Edsj__2017}%
  \BibitemOpen
  \bibfield  {author} {\bibinfo {author} {\bibfnamefont {J.}~\bibnamefont
  {Edsjo}}, \bibinfo {author} {\bibfnamefont {J.}~\bibnamefont {Elevant}},
  \bibinfo {author} {\bibfnamefont {R.}~\bibnamefont {Enberg}},\ and\ \bibinfo
  {author} {\bibfnamefont {C.}~\bibnamefont {Niblaeus}},\ }\bibfield  {title}
  {\bibinfo {title} {Neutrinos from cosmic ray interactions in the sun},\
  }\href {https://doi.org/10.1088/1475-7516/2017/06/033} {\bibfield  {journal}
  {\bibinfo  {journal} {Journal of Cosmology and Astroparticle Physics}\
  }\textbf {\bibinfo {volume} {2017}}\bibinfo  {number} { (06)},\ \bibinfo
  {pages} {033}}\BibitemShut {NoStop}%
\bibitem [{\citenamefont {Gutlein}\ \emph {et~al.}(2012)\citenamefont {Gutlein}
  \emph {et~al.}}]{Gutlein:2010ba}%
  \BibitemOpen
\bibfield  {number} {  }\bibfield  {author} {\bibinfo {author} {\bibfnamefont
  {A.}~\bibnamefont {Gutlein}} \emph {et~al.},\ }\bibfield  {title} {\bibinfo
  {title} {{Solar and Atmospheric Neutrinos: Limitations for Direct Dark Matter
  Searches}},\ }\bibfield  {booktitle} {\emph {\bibinfo {booktitle}
  {{Proceedings, 24th International Conference on Neutrino physics and
  astrophysics (Neutrino 2010): Athens, Greece, June 14-19, 2010}}},\ }\href
  {https://doi.org/10.1016/j.nuclphysbps.2012.09.173} {\bibfield  {journal}
  {\bibinfo  {journal} {Nucl. Phys. Proc. Supl.}\ }\textbf {\bibinfo {volume}
  {229-232}},\ \bibinfo {pages} {536} (\bibinfo {year} {2012})},\ \Eprint
  {https://arxiv.org/abs/1009.3815} {arXiv:1009.3815 [hep-ph]} \BibitemShut
  {NoStop}%
\bibitem [{\citenamefont {Gaisser}\ \emph {et~al.}(2014)\citenamefont
  {Gaisser}, \citenamefont {Jero}, \citenamefont {Karle},\ and\ \citenamefont
  {van Santen}}]{PhysRevD.90.023009}%
  \BibitemOpen
  \bibfield  {author} {\bibinfo {author} {\bibfnamefont {T.~K.}\ \bibnamefont
  {Gaisser}}, \bibinfo {author} {\bibfnamefont {K.}~\bibnamefont {Jero}},
  \bibinfo {author} {\bibfnamefont {A.}~\bibnamefont {Karle}},\ and\ \bibinfo
  {author} {\bibfnamefont {J.}~\bibnamefont {van Santen}},\ }\bibfield  {title}
  {\bibinfo {title} {Generalized self-veto probability for atmospheric
  neutrinos},\ }\href {https://doi.org/10.1103/PhysRevD.90.023009} {\bibfield
  {journal} {\bibinfo  {journal} {Phys. Rev. D}\ }\textbf {\bibinfo {volume}
  {90}},\ \bibinfo {pages} {023009} (\bibinfo {year} {2014})}\BibitemShut
  {NoStop}%
\bibitem [{\citenamefont {Arg\"{u}elles}\ \emph {et~al.}(2018)\citenamefont
  {Arg\"{u}elles}, \citenamefont {Palomares-Ruiz}, \citenamefont {Schneider},
  \citenamefont {Wille},\ and\ \citenamefont {Yuan}}]{Arguelles_2018}%
  \BibitemOpen
  \bibfield  {author} {\bibinfo {author} {\bibfnamefont {C.~A.}\ \bibnamefont
  {Arg\"{u}elles}}, \bibinfo {author} {\bibfnamefont {S.}~\bibnamefont
  {Palomares-Ruiz}}, \bibinfo {author} {\bibfnamefont {A.}~\bibnamefont
  {Schneider}}, \bibinfo {author} {\bibfnamefont {L.}~\bibnamefont {Wille}},\
  and\ \bibinfo {author} {\bibfnamefont {T.}~\bibnamefont {Yuan}},\ }\bibfield
  {title} {\bibinfo {title} {Unified atmospheric neutrino passing fractions for
  large-scale neutrino telescopes},\ }\href
  {https://doi.org/10.1088/1475-7516/2018/07/047} {\bibfield  {journal}
  {\bibinfo  {journal} {Journal of Cosmology and Astroparticle Physics}\
  }\textbf {\bibinfo {volume} {2018}}\bibinfo  {number} { (07)},\ \bibinfo
  {pages} {047}}\BibitemShut {NoStop}%
\bibitem [{\citenamefont {Dziewonski}\ and\ \citenamefont
  {Anderson}(1981)}]{Dziewonski:1981xy}%
  \BibitemOpen
\bibfield  {number} {  }\bibfield  {author} {\bibinfo {author} {\bibfnamefont
  {A.~M.}\ \bibnamefont {Dziewonski}}\ and\ \bibinfo {author} {\bibfnamefont
  {D.~L.}\ \bibnamefont {Anderson}},\ }\bibfield  {title} {\bibinfo {title}
  {{Preliminary reference earth model}},\ }\href
  {https://doi.org/10.1016/0031-9201(81)90046-7} {\bibfield  {journal}
  {\bibinfo  {journal} {Phys. Earth Planet. Interiors}\ }\textbf {\bibinfo
  {volume} {25}},\ \bibinfo {pages} {297} (\bibinfo {year} {1981})}\BibitemShut
  {NoStop}%
\bibitem [{\citenamefont {Fedynitch}\ \emph {et~al.}(2017)\citenamefont
  {Fedynitch}, \citenamefont {Dembinski}, \citenamefont {Engel}, \citenamefont
  {Gaisser}, \citenamefont {Riehn},\ and\ \citenamefont
  {Stanev}}]{Fedynitch:2017M4}%
  \BibitemOpen
  \bibfield  {author} {\bibinfo {author} {\bibfnamefont {A.}~\bibnamefont
  {Fedynitch}}, \bibinfo {author} {\bibfnamefont {H.~P.}\ \bibnamefont
  {Dembinski}}, \bibinfo {author} {\bibfnamefont {R.}~\bibnamefont {Engel}},
  \bibinfo {author} {\bibfnamefont {T.}~\bibnamefont {Gaisser}}, \bibinfo
  {author} {\bibfnamefont {F.}~\bibnamefont {Riehn}},\ and\ \bibinfo {author}
  {\bibfnamefont {T.}~\bibnamefont {Stanev}},\ }\bibfield  {title} {\bibinfo
  {title} {{A state-of-the-art calculation of atmospheric lepton fluxes}},\
  }\href {https://doi.org/10.22323/1.301.1019} {\bibfield  {journal} {\bibinfo
  {journal} {PoS}\ }\textbf {\bibinfo {volume} {ICRC2017}},\ \bibinfo {pages}
  {1019} (\bibinfo {year} {2017})}\BibitemShut {NoStop}%
\bibitem [{\citenamefont {Aartsen}\ \emph {et~al.}(2020)\citenamefont {Aartsen}
  \emph {et~al.}}]{PhysRevLett.125.141801}%
  \BibitemOpen
  \bibfield  {author} {\bibinfo {author} {\bibfnamefont {M.~G.}\ \bibnamefont
  {Aartsen}} \emph {et~al.} (\bibinfo {collaboration} {IceCube
  Collaboration}),\ }\bibfield  {title} {\bibinfo {title} {{eV-Scale Sterile
  Neutrino Search Using Eight Years of Atmospheric Muon Neutrino Data from the
  IceCube Neutrino Observatory}},\ }\href
  {https://doi.org/10.1103/PhysRevLett.125.141801} {\bibfield  {journal}
  {\bibinfo  {journal} {Phys. Rev. Lett.}\ }\textbf {\bibinfo {volume} {125}},\
  \bibinfo {pages} {141801} (\bibinfo {year} {2020})}\BibitemShut {NoStop}%
\bibitem [{\citenamefont {Esteban}\ \emph {et~al.}(2019)\citenamefont
  {Esteban}, \citenamefont {Gonzalez-Garcia}, \citenamefont
  {Hernandez-Cabezudo}, \citenamefont {Maltoni},\ and\ \citenamefont
  {Schwetz}}]{NuFit41}%
  \BibitemOpen
  \bibfield  {author} {\bibinfo {author} {\bibfnamefont {I.}~\bibnamefont
  {Esteban}}, \bibinfo {author} {\bibfnamefont {M.~C.}\ \bibnamefont
  {Gonzalez-Garcia}}, \bibinfo {author} {\bibfnamefont {A.}~\bibnamefont
  {Hernandez-Cabezudo}}, \bibinfo {author} {\bibfnamefont {M.}~\bibnamefont
  {Maltoni}},\ and\ \bibinfo {author} {\bibfnamefont {T.}~\bibnamefont
  {Schwetz}},\ }\bibfield  {title} {\bibinfo {title} {Global analysis of
  three-flavour neutrino oscillations: synergies and tensions in the
  determination of $\theta_{23}$, $\delta_{CP}$, and the mass ordering},\
  }\href {https://doi.org/10.1007/JHEP01(2019)106} {\bibfield  {journal}
  {\bibinfo  {journal} {Journal of High Energy Physics}\ }\textbf {\bibinfo
  {volume} {2019}},\ \bibinfo {pages} {106} (\bibinfo {year}
  {2019})}\BibitemShut {NoStop}%
\bibitem [{\citenamefont {Ayres}\ \emph {et~al.}(2007)\citenamefont {Ayres}
  \emph {et~al.}}]{Ayres:2007tu}%
  \BibitemOpen
  \bibfield  {author} {\bibinfo {author} {\bibfnamefont {D.~S.}\ \bibnamefont
  {Ayres}} \emph {et~al.} (\bibinfo {collaboration} {NOvA}),\ }\bibfield
  {title} {\bibinfo {title} {{The NOvA Technical Design Report}}\ }\href
  {https://doi.org/10.2172/935497} {10.2172/935497} (\bibinfo {year}
  {2007})\BibitemShut {NoStop}%
\bibitem [{\citenamefont {Abe}\ \emph {et~al.}(2020)\citenamefont {Abe} \emph
  {et~al.}}]{T2K_CPPhase_Nature}%
  \BibitemOpen
  \bibfield  {author} {\bibinfo {author} {\bibfnamefont {K.}~\bibnamefont
  {Abe}} \emph {et~al.} (\bibinfo {collaboration} {The T2K Collaboration}),\
  }\bibfield  {title} {\bibinfo {title} {Constraint on the matter--antimatter
  symmetry-violating phase in neutrino oscillations},\ }\href@noop {}
  {\bibfield  {journal} {\bibinfo  {journal} {Nature}\ }\textbf {\bibinfo
  {volume} {580}},\ \bibinfo {pages} {339} (\bibinfo {year}
  {2020})}\BibitemShut {NoStop}%
\bibitem [{\citenamefont {Adamson}\ \emph {et~al.}(2017)\citenamefont {Adamson}
  \emph {et~al.}}]{PhysRevLett.118.231801}%
  \BibitemOpen
  \bibfield  {author} {\bibinfo {author} {\bibfnamefont {P.}~\bibnamefont
  {Adamson}} \emph {et~al.} (\bibinfo {collaboration} {NOvA Collaboration}),\
  }\bibfield  {title} {\bibinfo {title} {Constraints on oscillation parameters
  from ${\ensuremath{\nu}}_{e}$ appearance and
  ${\ensuremath{\nu}}_{\ensuremath{\mu}}$ disappearance in nova},\ }\href
  {https://doi.org/10.1103/PhysRevLett.118.231801} {\bibfield  {journal}
  {\bibinfo  {journal} {Phys. Rev. Lett.}\ }\textbf {\bibinfo {volume} {118}},\
  \bibinfo {pages} {231801} (\bibinfo {year} {2017})}\BibitemShut {NoStop}%
\bibitem [{\citenamefont {Ishihara}(2019)}]{Ishihara:2019aao}%
  \BibitemOpen
  \bibfield  {author} {\bibinfo {author} {\bibfnamefont {A.}~\bibnamefont
  {Ishihara}} (\bibinfo {collaboration} {IceCube}),\ }\bibfield  {title}
  {\bibinfo {title} {{The IceCube Upgrade -- Design and Science Goals}}\
  }(\bibinfo {year} {2019})\ \Eprint {https://arxiv.org/abs/1908.09441}
  {arXiv:1908.09441 [astro-ph.HE]} \BibitemShut {NoStop}%
\bibitem [{\citenamefont {Adrian-Martinez}\ \emph {et~al.}(2016)\citenamefont
  {Adrian-Martinez} \emph {et~al.}}]{Adrian-Martinez:2016fdl}%
  \BibitemOpen
  \bibfield  {author} {\bibinfo {author} {\bibfnamefont {S.}~\bibnamefont
  {Adrian-Martinez}} \emph {et~al.} (\bibinfo {collaboration} {KM3Net}),\
  }\bibfield  {title} {\bibinfo {title} {{Letter of intent for KM3NeT 2.0}},\
  }\href {https://doi.org/10.1088/0954-3899/43/8/084001} {\bibfield  {journal}
  {\bibinfo  {journal} {J. Phys.}\ }\textbf {\bibinfo {volume} {G43}},\
  \bibinfo {pages} {084001} (\bibinfo {year} {2016})},\ \Eprint
  {https://arxiv.org/abs/1601.07459} {arXiv:1601.07459 [astro-ph.IM]}
  \BibitemShut {NoStop}%
\bibitem [{\citenamefont {Abe}\ \emph {et~al.}(2011)\citenamefont {Abe} \emph
  {et~al.}}]{Abe:2011ts}%
  \BibitemOpen
  \bibfield  {author} {\bibinfo {author} {\bibfnamefont {K.}~\bibnamefont
  {Abe}} \emph {et~al.},\ }\bibfield  {title} {\bibinfo {title} {{Letter of
  Intent: The Hyper-Kamiokande Experiment --- Detector Design and Physics
  Potential ---}},\ }\href@noop {} {\  (\bibinfo {year} {2011})},\ \Eprint
  {https://arxiv.org/abs/1109.3262} {arXiv:1109.3262 [hep-ex]} \BibitemShut
  {NoStop}%
\end{thebibliography}%

\end{document}